# Quantum Prediction of Ultra-Low Thermal Conductivity in Lithium Intercalation Materials

Tianli Feng[1,2,*,a], Andrew O'Hara[1], Sokrates T. Pantelides[1,2,†]

[1]*Department of Physics and Astronomy and Department of Electrical Engineering and Computer Science, Vanderbilt University, Nashville, Tennessee 37235, USA*

[2]*Center for Nanophase Materials Sciences, Oak Ridge National Laboratory, Oak Ridge, Tennessee 37831, USA*

**Lithium-intercalated layered transition-metal oxides, $Li_xTMO_2$, brought about a paradigm change in rechargeable batteries in recent decades and show promise for use in memristors, a type of device for future neural computing and on-chip storage. Thermal transport properties, although being a crucial element in limiting the charging/discharging rate, package density, energy efficiency, and safety of batteries as well as the controllability and energy consumption of memristors, are poorly managed or even understood yet. Here, for the first time, we employ quantum calculations including high-order lattice anharmonicity and find that the thermal conductivity $\kappa$ of $Li_xTMO_2$ materials is significantly lower than hitherto believed. More specifically, the theoretical upper limit of $\kappa$ of $LiCoO_2$ is 6 W/m-K, 2-6 times lower than the prior theoretical predictions. Delithiation further reduces $\kappa$ by 40-70% for $LiCoO_2$ and $LiNbO_2$. Grain boundaries, strains, and porosity are yet additional causes of thermal-conductivity reduction, while Li-ion diffusion and electrical transport are found to have only a minor effect on phonon thermal transport. The results elucidate several long-standing issues regarding the thermal transport in lithium-intercalated materials and provide guidance toward designing high-energy-density batteries and controllable memristors.**

*email: fengt@ornl.gov    †email: pantelides@vanderbilt.edu

[a] Present address: Energy and Transportation Science Division, Oak Ridge National Laboratory, Oak Ridge, Tennessee 37831, USA

# 1. Introduction

The discovery of layered lithium transition-metal oxides ($LiTMO_2$), especially $LiCoO_2$, had a major impact on rechargeable batteries in the past few decades[1,2]. $LiTMO_2$ stores lithium ions between $TMO_2$ layers with extraordinary mobility, energy density, power density, stability, and long cycle life. $LiTMO_2$-based $Li^+$-ion batteries are now widely used in most portable electronic devices[3–7]. Recently, the family of $LiTMO_2$ materials, including $LiCoO_2$ and $LiNbO_2$, were also found to be promising candidates for memristors[8–11], a type of device that can retain internal resistance based on the history of applied voltage and current, with potential for neural computing and on-chip storage[12–14]. However, it was realized in recent years that boosting the performance of batteries and memristors is not a trivial task because performance is determined by many coupled properties[3,4,15–18]. Thermal transport properties are a crucial element in this mix as it limits the charging/discharging rate, package density, energy efficiency, and safety of batteries in many applications, especially electric vehicles[19–23]. The ultra-low thermal conductivity[24] of $LiCoO_2$ is the largest drawback that contrasts sharply with other extraordinary electrical and chemical properties for battery applications. In particular, the recently identified localized hotspots[25] push the thermal issue from outside to inside the LiCoO$_2$ electrodes. Thermal transport plays a major role in the operation of memristors as well, since the on/off states are switched via ionic movements driven by nanoscale Joule heat[8–11]. Ultra-low thermal conductivity is necessary to confine the Joule heat for the realization of in-memory computing with low energy consumption.

Despite their significance in various applications, the thermal transport properties of LiTMO$_2$ have not been adequately explored or understood compared to electrical and ionic transport properties, due to limitations of theoretical and experimental capabilities. Theoretically, the evaluation of high-order lattice anharmonicity and phonon scattering by parameter-free quantum calculations has until recently[26–28] been inaccessible. Available theoretical studies have been based on either classical molecular dynamics[29] or lowest-order anharmonicity[30], which lack sufficient predictive power. As shown in Ref.[31], high-order lattice anharmonicity can induce high-order phonon-phonon scattering, which provides additional phonon-scattering channels and increases the total phonon scattering rates or, equivalently, reduces the phonon relaxation times and thermal conductivity. It was shown that high-order phonon-phonon scattering is generally important for low thermal conductivity materials, complex crystals with many phonon branches, materials with ionic Coulombic interactions, and most materials at high temperatures. At the same time, experimental work has been limited to polycrystals[24]. Overall, the available theoretical and experimental works have not provided deep physical insights in the underlying quantum mechanisms. As a result, several fundamental yet practically significant questions have remained open:

*(1) What are the intrinsic (anisotropic) thermal conductivities of Li$_x$TMO$_2$?*

*(2) What is the impact of charging on the vibrational properties and thermal conductivity of Li$_x$TMO$_2$ (x≤1)?*

*(3) As most practical devices are made using polycrystals, how strongly does the grain size affect the thermal conductivity of $Li_xTMO_2$ polycrystals? How much can we tune the thermal conductivity by tuning the grain size via tuning the annealing temperature?*

*(4) As large internal strains are usually generated during charging/discharging cycles, what is the impact of strain on the thermal transport?*

*(5) As lithium atoms are mobile, making $Li_xTMO_2$ half-solid half-liquid-like, what is the impact of Li liquidity on the thermal conductivity?*

In this paper, for the first time, we fully address these questions by high-throughput quantum mechanical calculations including up to fourth-order anharmonicity and four-phonon scattering, considering $Li_xCoO_2$ (x = 1, 0.67, 0.5, 0.33) and $Li_xNbO_2$ (x = 1, 0.5) as prototype examples. We predict that the thermal conductivities are significantly lower than hitherto believed. The origins of the ultra-low thermal conductivities are as follows:

(1) The fourth-order anharmonicity in these materials is strong and the impact of four-phonon scattering is significant. As a result, the intrinsic thermal conductivity is much smaller than has been predicted without inclusion of these phenomena. The intrinsic in-plane and through-plane thermal conductivities are 9.7 and 1.4 $Wm^{-1}K^{-1}$ for $LiCoO_2$, in contrast to previously predicted[29,30,32,33] 19.8 - 53.6 $Wm^{-1}K^{-1}$ and 2.2 - 8.4 $Wm^{-1}K^{-1}$, respectively. In comparison to $LiCoO_2$, weaker anharmonicity is present in $LiNbO_2$ because Nb has fewer lone electrons than Co.

(2) Charging (delithiation) is another significant effect that reduces $\kappa$. It strongly softens the lattice, reduces phonon velocities (especially in the through-plane direction), increases the anisotropic ratio, and increases anharmonicity as well the three- and four-phonon scattering rates in all $Li_xTMO_2$ studied in this work. The calculated $\kappa$ matches available experimental data only when four-phonon scattering is included, supporting the significance of high-order anharmonicities as well as the validity of the present calculations.

(3) Grain size is another factor that is responsible for the ultra-low $\kappa$ of LiTMO$_2$ polycrystals. Although four-phonon scattering reduces the phonon mean free paths and thus weakens the grain-boundary effect, $\kappa$ of LiCoO$_2$ and LiNbO$_2$ can still be reduced by 26% and 60%, respectively, when grain size is reduced down to 10 nm.

(4) Internal strains significantly tune the thermal conductivity by changing the phonon scattering phase space. 2% compressive strain increases the thermal conductivity by 30-80% (depending on the direction), while 2% tensile strain diminishes the thermal conductivity. Also, we find that the vibration frequency of Li is exceptionally sensitive to strain and may be used in detecting strains in experiments.

(5) The impact of Li diffusion on the thermal transport is negligible since the hopping rate is much smaller than phonon scattering rates. Other possible factors such as electronic thermal conductivity and porosity are also discussed. These results elucidate long-standing questions regarding thermal transport in Li$_x$TMO$_2$ and provide valuable insights that can guide the design of next-generation batteries and electronics.

## 2. Methods

A detailed description of the methods and formalism is given in the Supplemental Information. Here, we briefly introduce the methodology that we used. The thermal conductivity of single crystals is calculated using exact solutions of the Boltzmann transport equation in the phonon-relaxation-time approximation, including three- and four-phonon scattering. Density functional theory calculations are carried out to produce the 2$^{nd}$-, 3$^{rd}$-, and 4$^{th}$-order interatomic force constants. Fermi's golden rule is used to evaluate the three- and four-phonon scattering rates. After we obtained the intrinsic anisotropic thermal conductivities of the crystals, we use an orientation average to evaluate the thermal conductivity of polycrystals. To account for the impact of delithiation, we remove varying amounts of Li atoms from the unit cells and recalculate all the force constants and phonon scattering rates. We then add an extra scattering term, accounting for phonon-Li disorder scattering, in addition to the intrinsic phonon-phonon scattering, to calculate the thermal conductivity. When the grain size is comparable to or even smaller than the phonon mean free path, the grain boundary could confine the phonons from propagating, and therefore reduce the phonons' contribution to the thermal conductivity

## 3. Results and Discussion

**3.1 Vibrational and thermal properties and the impact of delithiation**. We start by discussing the calculated vibrational properties of $LiCoO_2$ in the atomic structure shown in Fig. 1 **a**. Lithium atoms are intercalated in between $CoO_2$ layers. The through-plane arrangement of these layers is ABC structure, or O3 structure, based on the classification by Ref. [34]. The phonon dispersion and projected density of states (pDOS) are shown in Figs. 1 **b** and **g**. It is found that the Co, Li, and O species have distinct ranges of vibrational frequencies, i.e., they dominate the low-, mid-, and high-frequency ranges, respectively. Co has the highest atomic mass and therefore vibrates most slowly. Li atoms, although being the lightest, are bonded loosely in the lattice, and therefore vibrate more slowly than O atoms, which bond strongly in the lattice with Co atoms and vibrate most rapidly.

To examine the effects of gradual charging (removing Li from $Li_xCoO_2$), we calculated the phonon dispersions and pDOS of $Li_xCoO_2$ for x=0.67, 0.5, and 0.33. As shown in Supplementary Figs. S1 and S2, the $Li_xTMO_2$ crystalline structures structures at certain x values are obtained by removing Li atoms periodically and keeping the smallest possible unit cell after Li removal. These structures are all obtained from Materials Projects [35]. For example, in $Li_{0.67}CoO_2$, the remaining Li atoms form a graphene-like structure with the center Li in each hexagonal ring removed. In $Li_{0.33}CoO_2$, the Li removal is reversed, i.e., the center Li in each hexagonal ring is kept while all Li atoms on hexagonal rings are removed. In $Li_{0.5}CoO_2$, Li is removed from every other row in each layer, leaving half the number of Li atoms remaining in the lattice. $LiNbO_2$ has a similar layered intercalation structure with $LiCoO_2$ but with a different stacking order of ABAB. The Li atoms in all layers are overlapped in the top view. The creation of crystalline delithiated $Li_xNbO_2$ structures is similar to that of $Li_xCoO_2$. The phonon dispersion and harmonicity are calculated based on the crystalline $Li_xTMO_2$ structures, while the lithium disorder is taken into account by including an extra Li disorder phonon-scattering term in the thermal conductivity calculations. The results, shown in Figs. 1 **c-e,** respectively, reveal an interesting phenomenon: the lower phonon branches of $LiCoO_2$ gradually move to even lower energies (they soften) while the upper branches move to even higher energies (they stiffen), generating a gap between them. We further track these changes in Figs. 1 **h-j**, where we compare the phonon pDOS of the Li, Co, and O species and relate the changes to the atomic structures (more comparisons are included in Supplementary Figs. S3-S5). We find that the softening of lower branches is effected by Co and Li while the stiffening of upper branches is dominated by Co and O. When the Li concentration is sufficiently high, Li atoms act as linkages between adjacent $CoO_2$ layers by attracting $O^{2-}$ ions. We conclude that Li loss leads to weakening of the interlayer bonding strength and hence softening of the lower phonon branches, which describe the in-phase vibrations of the lattice. These features also explain the softening of the bulk modulus during delithiation observed in experiments[24]. In contrast, interlayer softening is accompanied by intralayer stiffening, i.e., the binding between Co and O becomes stronger. This strengthening arises because the valence of Co increases from +3 to +3+(1-x) after delithiation, which strengthens the Coulombic interaction between $Co^{3+1-x}$ and $O^{2-}$ ions. As a result, the upper phonon branches move to even higher energies because they are dominated by out-of-phase vibrations between the $Co^{3+1-x}$ and $O^{2-}$ ions. In addition, we note that, after delithiation, a flat band at around 12 THz lies between the lower and upper branches. The eigenvector shows that this mode is the through-plane vibration of Li atoms, as can be seen in the pDOS of Li in Figs. 1 **h-j**. This feature indicates that Li is loosely bonded in the lattice and can rattle between the layers. All these characteristics found in vibrational properties provide physical insights in the thermal transport as will be discussed later in this paper.

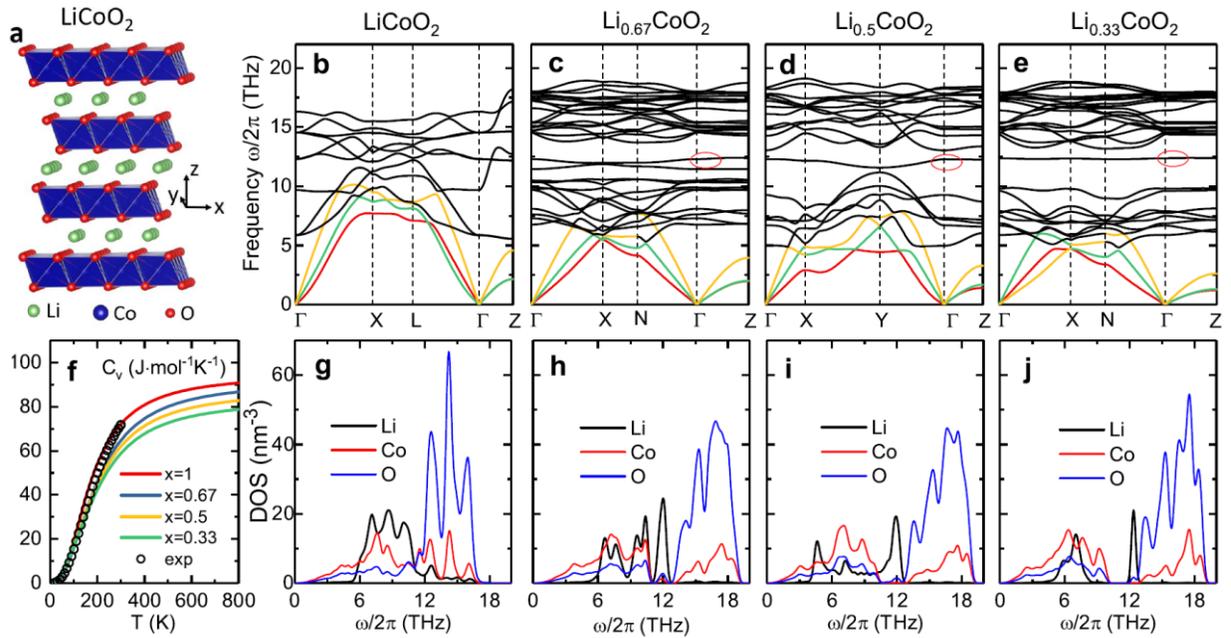

**Figure 1 | Harmonic vibrational properties of $Li_xCoO_2$ (x=1, 0.67, 0.5, 0.33). a**, Crystal structure of $LiCoO_2$. **b-e**, Phonon dispersion relations of $Li_xCoO_2$. The colored curves highlight acoustic branches. The red circles in **c-e** highlight the flat band dominated by Li vibrations. **f**, Heat capacities of $Li_xCoO_2$ as functions of temperature, compared to experimental data[36]. **g-j**, Atomic-site-projected phonon densities of states.

The impact of delithiation on phonon velocities is also significant. Generally, the group velocities of most phonon modes are significantly reduced as the lattice becomes softer during delithiation (Supplementary Figs. S6-S9). Especially, the velocities of the phonons associated with through-plane vibrations, regardless of transport direction, are softened. For example, the through-plane long-wavelength longitudinal-acoustic ($v_{\perp,LA}$) and transverse-acoustic ($v_{\perp,TA}$) phonon velocities decrease from 5.6 and 3.0 km/s to 3.7 and 1.9 km/s, respectively, when $x$ decreases from 1 to 0.33 in $Li_xCoO_2$. Most in-plane phonon velocities are reduced as well, except for the long-wavelength (in-phase) $v_{\perp,LA}$ and $v_{\perp,TA}$, which increase with delithiation due to the in-plane stiffening discussed above. This result is analogous to the phonon-focusing effect[37], by which compression of a lattice in one direction reduces the phonon velocity in the perpendicular direction since phonon wavevectors gain larger projection along the compressed direction.

The heat capacity $C_v$ of $Li_xCoO_2$, which is also determined by the vibrational spectra, is also strongly affected by delithiation. As shown in Fig. 1 **f**, $C_v$ increases with temperature and saturates above 800 K, agreeing well with available experimental data[36]. Upon delithiation, the room-temperature (RT) $C_v$ decreases almost linearly with decreasing Li concentration at a rate of 0.2 J mol$^{-1}$ K$^{-1}$ per 1at%Li (Supplementary Fig. S10). Full delithiation can lead to 28% loss of heat capacity, which indicates a faster temperature rise at low Li concentrations, given the same amount of Joule heat.

Although so far we discussed the vibrational properties of $Li_xCoO_2$, the characteristics and the impact of delithiation are equally applicable to $Li_xNbO_2$ (Supplementary Figs. S11-S15), possibly indicating general trends in layered $Li_xTMO_2$.

**3.2 Third and fourth-order anharmonicities and phonon scattering rates**. Lattice anharmonicity determines the thermal resistance of a system. Previously, most studies were based on the lowest order anharmonicity, i.e., third-order anharmonicity and three-phonon scattering[38–40]. Beginning in 2016, Feng et al.[26–28], computed fourth-order anharmonicity and four-phonon scattering and found that their effect on thermal transport can be significant. However, the capability of those four-phonon studies was limited to simple systems with two basis atoms per unit cell. The 4-14 basis atoms per unit cell in the $Li_xTMO_2$ systems make the four-phonon scattering calculations extremely demanding. Here, we calculate the third-order anharmonicity and three-phonon scattering for all the materials, and then take $LiCoO_2$, $Li_{0.33}CoO_2$, $LiNbO_2$ and $Li_{0.5}NbO_2$ to explore the significance of fourth-order anharmonicity and four-phonon scattering by using high-throughput calculations.

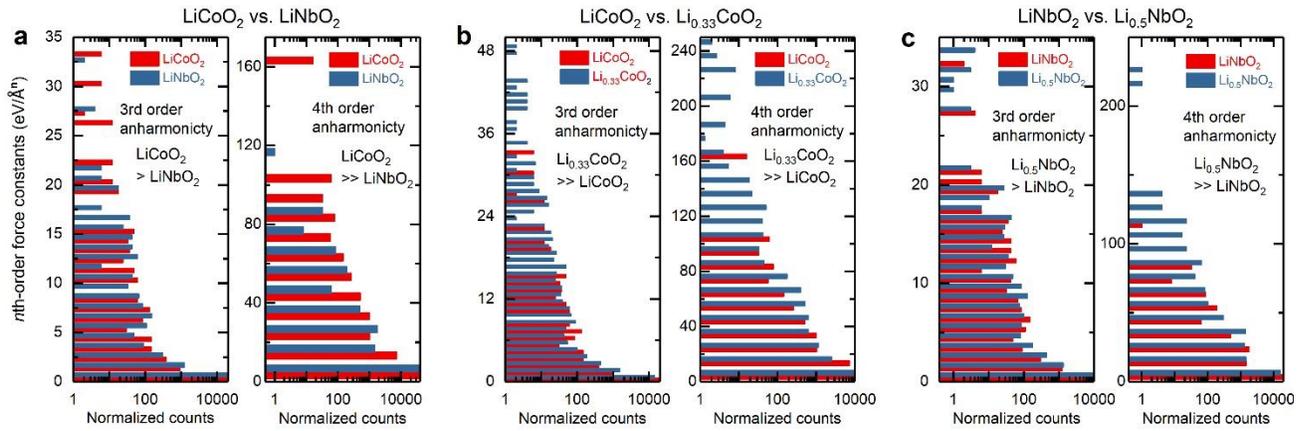

**Figure 2 | Third and fourth-order anharmonicities of $Li_xCoO_2$ and $Li_xNbO_2$. a**, Comparison between $LiCoO_2$ and $LiNbO_2$. **b**, Comparison between $LiCoO_2$ and $Li_{0.33}CoO_2$. **c**, Comparison between $LiNbO_2$ and $Li_{0.5}NbO_2$. The subfigures count the number of force-constant values in certain ranges. The counts of each material are normalized by the number of $TMO_2$ in a unit cell.

The comparison of the anharmonicities between two materials is done by comparing their largest values of anharmonic force constants (AFCs) and their counts (when AFCs values are close). See Supplementary Section 2 for details of the counting mechanism of AFCs and the rationale of qualitative or quantitative interpretation. Compared to the previous method of qualitatively comparing certain anharmonic vibrational potential wells[41], the present method is more accurate as it quantifies all significant anharmonic potential wells in the systems. Figure 2 shows the third- and fourth-order AFCs for **a**, $LiCoO_2$ and $LiNbO_2$, **b**, $LiCoO_2$ and $Li_{0.33}CoO_2$, and **c**, $LiNbO_2$ and $Li_{0.5}NbO_2$. Several representative relations between anharmonicities and coordinates, TM species, as well as Li concentrations are revealed:

(1) Most of the largest AFCs are associated with the through-plane vibrations.

(2) Most of the significant AFCs are associated with TM or O atoms, rather than Li atoms.

(3) $LiCoO_2$ is more anharmonic than $LiNbO_2$ in the third order and is much more so in the fourth order, which may originate from the fact that Co ($4s^2 3d^7$) has more lone electrons than Nb ($5s^1 4d^4$) after binding with O. These lone electrons serve as a buffer that softens the vibrations[42].

(4) $Li_{0.33}CoO_2$ and $Li_{0.5}NbO_2$ are much more anharmonic than $LiCoO_2$ and $LiNbO_2$ in both third and fourth orders, demonstrating the increase of anharmonicity with delithiation.

To sum up, the above comparisons reveal that the interaction between TM and O becomes more anharmonic with the loss of Li atoms in $LiCoO_2$, $LiNbO_2$, and possibly other layered $Li_xTMO_2$, which provides valuable insight for understanding the thermal transport.

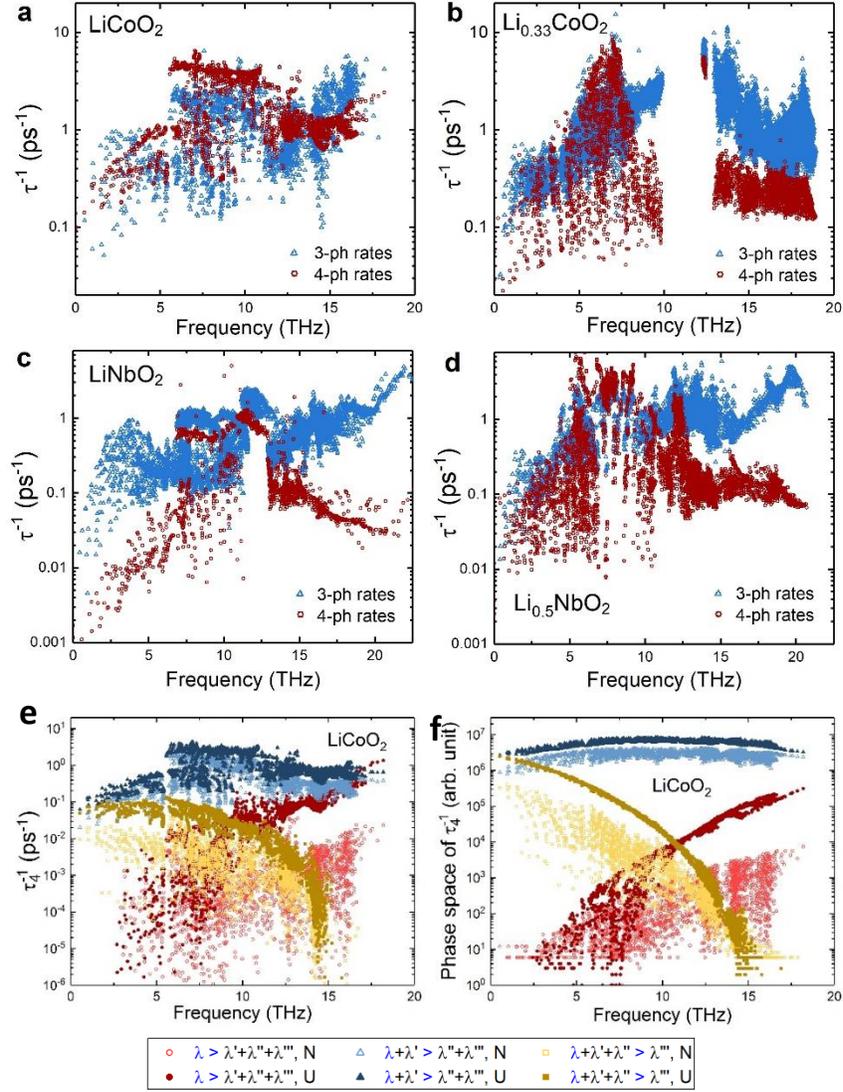

**Figure 3 | Full-Brillouin zone anharmonic three- and four-phonon scattering rates in $Li_xCoO_2$ and $Li_xNbO_2$ at room temperature. a-d**, Three and four-phonon scattering rates in $LiCoO_2$, $Li_{0.33}CoO_2$, $LiNbO_2$, and $Li_{0.5}NbO_2$, respectively. **e**, Decomposed four-phonon scattering rates in $LiCoO_2$. **f**. Decomposed numbers of possible four-phonon scattering events in $LiCoO_2$. In **e** and **f**, the normal processes and Umklapp processes are represented by open and solid symbols, respectively. The splitting ($\lambda \to \lambda' + \lambda'' + \lambda'''$), redistribution ($\lambda + \lambda' \to \lambda'' + \lambda'''$), and recombination ($\lambda + \lambda' + \lambda'' \to \lambda'''$) processes are represented by red, blue and yellow colors, respectively.

With third and fourth order AFCs, the three and four-phonon scattering rates of $Li_xCoO_2$ and $Li_xNbO_2$ are calculated. Distinctly, as shown in Fig. 3 **a-d**, we find that four-phonon rates are substantial for most of the materials of interest here, providing a critical revisit of prevailing knowledge[30]. Among the four materials, $LiNbO_2$ is the least anharmonic and has the lowest three- and four-phonon scattering rates, consistent with the conclusions drawn from the analyses of AFCs. Compared to the well-known three-phonon scattering, four-phonon scattering has never been explored in complex crystals yet and is of particular interest. To gain deeper insight, we decompose the overall four-phonon scattering into the *splitting* ($\lambda \to \lambda' + \lambda'' + \lambda''' + R$), *redistribution* ($\lambda + \lambda' \to \lambda'' + \lambda''' + R$), and *recombination* ($\lambda + \lambda' + \lambda'' \to \lambda''' + R$) categories and distinguish the Normal ($R = 0$) from Umklapp processes ($R \neq 0$), where $\lambda = (q, j)$, a shorthand for a phonon mode with wavevector $q$ and dispersion branch $j$. $R$ can be any reciprocal lattice vector. The decomposed four-phonon scattering rates in $LiCoO_2$ are shown in Fig. 3 **e** (similar plots for $Li_{0.33}CoO_2$, $LiNbO_2$, and $Li_{0.5}NbO_2$ can be found in Supplementary Fig. S16). Clearly, four-phonon scattering is dominated by the Umklapp *redistribution* processes in all four materials. Taking a further step, we find that this is due to the large number of available combinations of ($\lambda', \lambda'', \lambda'''$) that satisfy both energy and momentum conservation rules for the $\lambda + \lambda' \to \lambda'' + \lambda''' + R$ process, as shown in Fig. 3 **f**. These processes are most intensive in the mid-frequency range. We also note another significant phenomenon, i.e., the four-phonon scattering rates of upper optical branches are not as high as expected in the literature on simple crystals[26,27], which is important for understanding the optical properties of complex materials.

**3.3 Lattice thermal conductivity**. The lattice thermal conductivity of $LiCoO_2$ single crystal calculated by first principles is shown in Fig. 4 **a**. Four-phonon scattering reduces the room-temperature intrinsic in-plane $\kappa$ ($\kappa_\parallel$) from 26.3 to 9.7 $Wm^{-1}K^{-1}$, and through-plane $\kappa$ ($\kappa_\perp$) from 3.8 to 1.4 Wm$^{-1}$K$^{-1}$, providing significant revision to the prevailing knowledge[29,30,32,33] which predicted 19.8 - 53.6 Wm$^{-1}$K$^{-1}$ and 2.2-8.4 Wm$^{-1}$K$^{-1}$ for $\kappa_\parallel$ and $\kappa_\perp$, respectively. The $\kappa$ reduction due to fourth-order anharmonicity is as large as 60% for both directions at room temperature, and increases to 83% as temperature climbs to 900 K.

The predicted $\kappa$ of Li$_x$CoO$_2$ after delithiation to x(Li)=0.33 is shown in Fig. 4 **b**. The $\kappa_{\parallel,34}$ and $\kappa_{\perp,34}$ are 6.5 and 0.41 Wm$^{-1}$K$^{-1}$, respectively, which are 33% and 70% smaller than that of LiCoO$_2$. Thus, we conclude that the loss of Li significantly affects the thermal transport, especially in the through-plane direction. In Li$_{0.33}$CoO$_2$, we find that the room-temperature $\kappa$ reduction due to fourth-order anharmonicity is 36%, which is smaller than that in LiCoO$_2$. This is because the importance of fourth-order anharmonicity is overwhelmed by the loss of Li in reducing $\kappa$.

Compared to Li$_x$CoO$_2$, Li$_x$NbO$_2$ has much larger thermal conductivity as shown in Figs. 4 **c-d**. $\kappa_\parallel$ and $\kappa_\perp$ are 31.5 and 5 Wm$^{-1}$K$^{-1}$ for LiNbO$_2$, and 10.6 and 1.1 Wm$^{-1}$K$^{-1}$ for Li$_{0.5}$NbO$_2$ single crystals, respectively. The effect of four-phonon scattering is smaller compare to Li$_x$CoO$_2$ but still non-negligible. However, the $\kappa$ reduction due to delithiation is substantially stronger: delithiation to x(Li)=0.5 leads to a reduction of 66% and 78% for $\kappa_\parallel$ and $\kappa_\perp$, respectively. Such significant reduction may pose potential challenge for Li$_x$NbO$_2$ batteries[7], but gives promise for Li$_x$NbO$_2$ thermoelectrics[43]. We also note that the $\kappa$ anisotropic ratio for Li$_x$CoO$_2$, Li$_x$NbO$_2$, and possibly other layered Li$_x$TMO$_2$ materials is large and increases with delithiation, which may provide promise for the further new-directional thermal management.

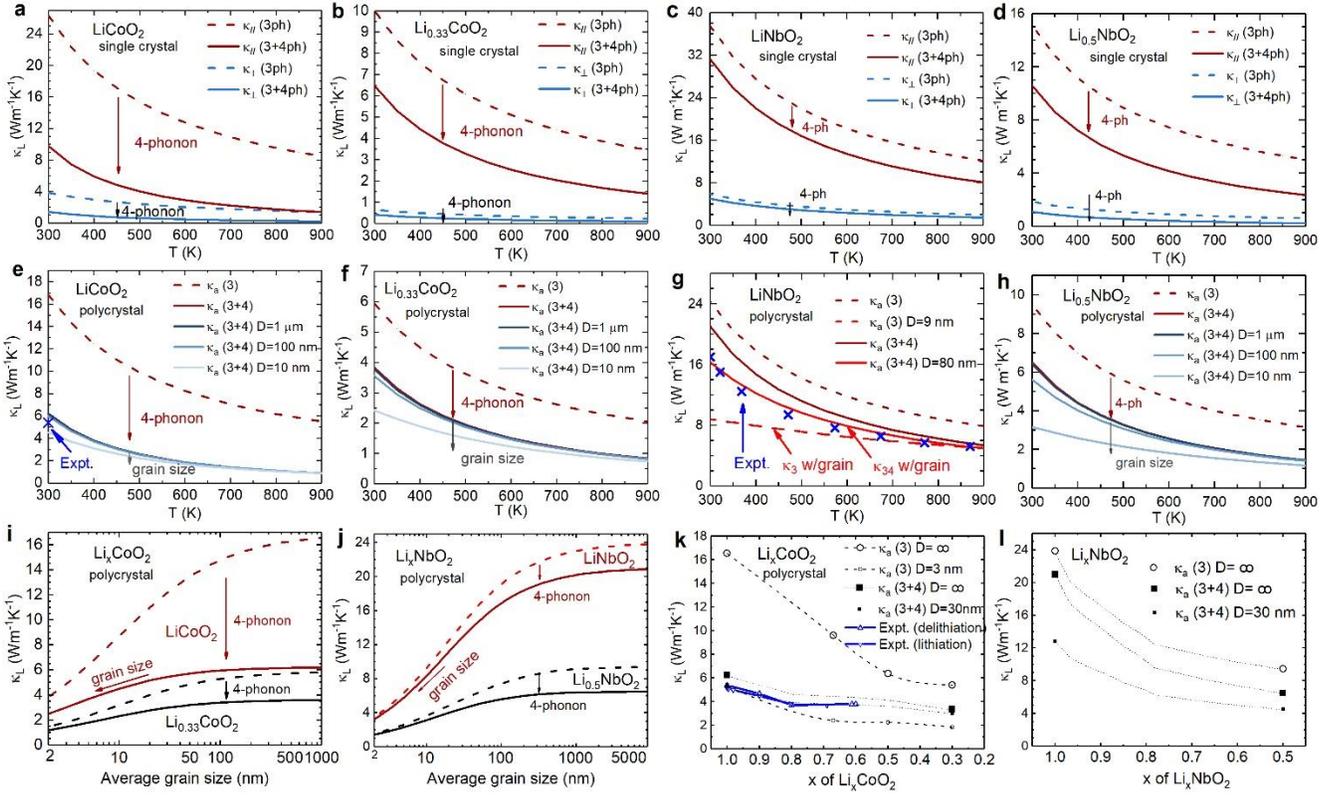

**Figure 4 | Lattice thermal conductivities $\kappa_L$ of $Li_xCoO_2$ and $Li_xNbO_2$ predicted by first principles.** In all the panels, dashed and solid curves represent the predictions with and without four-phonon scattering, respectively. **a-d**, In-plane ($//$) and through-plane ($\perp$) thermal conductivities of $LiCoO_2$, $Li_{0.33}CoO_2$, $LiNbO_2$, and $Li_{0.5}NbO_2$ single crystals as functions of temperature. **e-f**, Orientation-averaged thermal conductivities ($\kappa_a$) of $LiCoO_2$, $Li_{0.33}CoO_2$, $LiNbO_2$, and $Li_{0.5}NbO_2$ polycrystals as functions of temperature. The predictions with various grain sizes ($D$) are also included. **i,j**, Room-temperature polycrystalline thermal conductivities, $\kappa_a(300K)$, as functions of grain diameter ($D$). **h,k**, $\kappa_a(300K)$ as functions of Li concentration $x$. In **h** and **k**, the dotted curves connecting the four-phonon predictions are drawn to guide eyes. Experimental data of $Li_xCoO_2$ and $LiNbO_2$ are taken from Ref.[24] and Ref.[43], respectively.

To compare with available experimental data[24,43] and validate the present calculation of four-phonon scattering, we convert the calculated anisotropic $\kappa_{\parallel,\perp}$ of single crystals to the orientation-average $\kappa_a$ of polycrystals (see Supplementary Section 3 for details). The $\kappa$ without and with four-phonon scattering are represented by $\kappa_a(3)$ and $\kappa_a(3+4)$, respectively. Clearly, $\kappa_a$ strongly depends on the temperature, grain size $D$, and lithium concentration $x$, as shown in Figs. 4 **e-h**, **i-j**, and **k-l**, respectively. Since the calculated $\kappa_a(3)$ and $\kappa_a(3+4)$ depend on $D$, which was not determined in experimental works[24,43], it is not reasonable to validate $\kappa_a(3)$ or $\kappa_a(3+4)$ by doing a single $\kappa$ value comparison since $D$ is an adjustable fitting parameter, which can be adjusted arbitrarily to match experimental $\kappa_{\text{exp}}$. However, we can validate the necessity and accuracy of four-phonon scattering calculations by comparing the $T$ dependency since $\kappa_a(3+4)$ decreases faster than $\kappa_a(3)$ with $T$, especially when defects or grain boundaries are presented[26–28]. For example, as shown in Fig. 4 **g** by the red dashed line, the experimental data cannot be fitted without inclusion of four-phonon scattering. Although a $D$ value can be chosen for $\kappa_a(3)$ to match $\kappa_{\text{exp}}$ at a single temperature (900 K, for example), this $D$ value results in disagreement at all the other temperatures. In contrast, $\kappa_a(3+4)$ can match the experimental data throughout the whole temperature range with an appropriate $D$ value. Similarly, the validation can also be done by comparing the lithium concentration ($x$) dependency, as shown in Fig. 4 **k**. That is, although a $D$ value can be chosen for $\kappa_a(3)$ to match $\kappa_{\text{exp}}$ at a single $x$ ($x$=0, for example), this $D$ value results in disagreement at all the other $x$ values, while $\kappa_a(3+4)$ can match with experimental data throughout the whole $x$ range with an appropriate $D$ value. In addition, in comparison to the $D$=3 nm fitted by using $\kappa_a(3)$, the fitted $D$ by using $\kappa_a(3+4)$ is about 30 nm, which agrees better with the 20-50 nm shown in the high-resolution transmission-electron-microscopy images[24]. To sum up, by validating the predicted $\kappa_a(3+4)$ with available experimental data[24,43] through different perspectives for different materials with broad temperature and lithium concentration ranges, it is safe to conclude the significance of four-phonon scattering in $Li_xCoO_2$ and $Li_xNbO_2$ and the accuracy of the present work.

To gain more insight into the thermal transport, we investigate the spectral thermal conductivity contributions in these materials. It is found that $\kappa_\parallel$ of $LiCoO_2$ and $Li_{0.33}CoO_2$ are contributed by the phonons with frequencies 0-10 THz and 0-5 THz, respectively. The $\kappa_\perp$ comes from 0-5 THz and 0-3 THz, respectively (Supplementary Figs. S17). These frequency ranges have little overlap with the Li pDOS, indicating that the vibrations of Li ions are localized and do not contribute to the thermal transport. The results also indicate that, with delithiation, the lower-branch acoustic phonons are more dominant in thermal transport. This dominance of the lower-branch acoustic phonons occurs because the higher phonon branches (dominated by strong interactions between $Co^{4-x}$ and $O^{2-}$) become more anharmonic as discussed in the preceding text. Similar conclusions are also applied to $Li_xNbO_2$ (Supplementary Figs. S18). The phonon relaxation times of the lower branches are 1~2 orders of magnitude longer than the upper branches (Supplementary Figs. S19-S26).

**3.4 Significant strain effect**. During the charging/discharging process, large strains can be accumulated[5,44,45], with $Li_xCoO_2$ being able to sustain as large as 30% through-plane strain[46]. The strain impact on the mechanical, electrical, and ionic transport properties have been extensively studied[46–49], while the impact on thermal transport is still unclear. Taking delithiated $Li_xCoO_2$ (x=0.33) for example, we investigated the strain effect on the harmonic and third-order anharmonic properties under 2% (tensile), -2% (compressive), -5% (compressive), and -10% (compressive) strains along the $z$ direction. We find that, remarkably, apart from the slight softening or stiffening effect due to the tensile or compressive strains, there is an exceptional variation in the vibrations of Li atoms under strain (Supplementary Figs. S27-S32). As shown in Fig. 5 **a**, the Li vibrations shift significantly with the applied strain, with approximately a linear rate of 0.4 and 0.3 THz per 1% strain, i.e., 11 and 8 cm$^{-1}$ per GPa, for the two peak frequencies in the pDOS. The shifting rates are more than 1 order of magnitude larger than those of other modes. The main reason is that the bonding of Li with the other atoms is very weak and highly dependent on the distance to the adjacent TMO$_2$ layers. A slight change in the interlayer spacing due to strain can therefore have a large effect on the vibrational frequency of Li. Inspired by the temperature measurement via temperature-dependent frequency shifts[50], we propose that the strong strain-dependent Li partial DOS provides a potential tool to measure the strain of Li$_x$CoO$_2$ in experiments. For example, by using either optical methods such as Raman and infrared spectroscopy or electron methods such as electron energy loss spectroscopy (EELS), the lithium phonon energy can be measured and then the average strain in the sample can be obtained. If an electron method is used, the measurement may also have high spatial resolution so that variations in the local strain of the material can be detected.

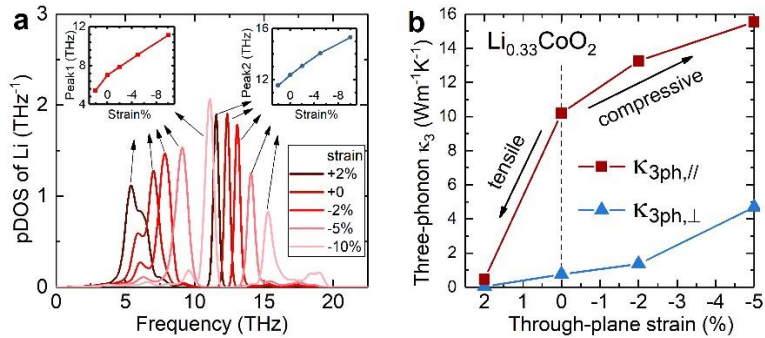

**Figure 5 | Exceptional strain impact on thermal properties of Li$_x$CoO$_2$. a**, Projected phonon DOS on Li in Li$_{0.33}$CoO$_2$ with +2%, 0, -2%, -5%, -10% through-plane strains. The insets show the shifts of the two peak frequencies as functions of strain. **b**, Room-temperature in-plane and through-plane three-phonon thermal conductivity of Li$_{0.33}$CoO$_2$ under various through-plane strains.

In addition to the remarkable change in harmonic vibrations, we find that strain has surprising impact on the thermal conductivity as well. As shown in Fig. 5 **b**, 2% tensile strain can nearly diminish the three-phonon $\kappa_3$ of $Li_{0.33}CoO_2$. 2% compressive strain can increase $\kappa_3$ by 30% and 80% for the $\kappa_\parallel$ and $\kappa_\perp$, respectively. The origin of such surprisingly large impact of strain is elucidated by comparing the three factors that determine $\kappa$: phonon velocities, anharmonicities, and scattering phase space. It turns out that the velocities and anharmonicities do not vary significantly with strain (Supplementary Figs. S33 and S34), while the scattering phase space depends strongly on strain due to the change in Li pDOS (Supplementary Fig. S35). With compressive strains, the phonon dispersion tends to spread out, and the energy spacing between different phonon branches enlarges. As a result, the phonon-phonon scattering probability decreases since the energy conservation rule is not as easily satisfied compared to more crowded branches. Therefore, the phonon scattering rate is reduced and the phonon relaxation time is increased, leading to a sharp increase of thermal conductivity. With tensile strains, the scenario is reversed. Note that these $\kappa_3$ values are not the exact thermal conductivities of materials since four-phonon scattering or *diffuson* contribution[51] is not included in the calculation of $\kappa_3$. Nevertheless, studying $\kappa_3$ is enough to reveal the general impact of strain on anharmonicity.

The calculated phonon scattering rates and intrinsic $\kappa$ in the preceding text can enable the estimation of the impact of other effects, including ionic diffusion, electrical conductivity, and porosity, on the thermal transport. These effects are of great interest in both fundamental physics and practical application in batteries and memristors. The thermal conductivity derived from vibrational properties assumes that the atoms vibrate around fixed positions, i.e., their equilibrium positions, and do not flow around. It has, therefore, been a persistent question whether the diffusion of atoms makes a difference to the thermal conductivity of the materials with high ionic mobility[52]. With detailed analysis shown in Supplementary Section 18, 19 and 20, we conclude that the impact of Li diffusion and electrical conductivity on the thermal properties of $Li_xCoO_2$ is negligible, and the impact of porosity is large in experimental cathodes since they are usually porous. This impact can be readily predicted by using the effective medium approximation.

The analysis in this work is mainly based on layered $Li_xTMO_2$; however, we expect that most conclusions of this study can also be applied to other Li-based TMOs such as $LiMn_2O_4$ (spinel structure) and $LiFePO_4$ (olivine structure). For example, most of these Li-intercalation materials are also likely to have strong high-order lattice anharmonicity and four-phonon scattering (due to the soft bonding of Li and the anharmonic Coulombic interaction between TM and O atoms), strong impact of delithiation, strong effect of grain boundaries, and negligible impact of Li fluidity. Nevertheless, the impact of strain is not necessarily as large as that in layered $Li_xTMO_2$ in which the through-layer strain can easily change the Li bonding environment.

## 4. Conclusions

To conclude, by considering up to the fourth order quantum lattice anharmonicity and four-phonon scattering, we have resolved several long-standing, fundamental, yet practically important questions regarding the vibrational and thermal properties of $LiCoO_2$, $LiNbO_2$ and possibly other layered $Li_xTMO_2$ that are crucial for battery and memristors applications. In particular, we showed that (1) the intrinsic thermal conductivities of $LiCoO_2$ and $LiNbO_2$ are 2-6 times lower than those predicted by prior theories due to strong four-phonon scattering; (2) although Li vibrations do not contribute to the thermal conductivity, the delithiation process can result in significant decrease of thermal conductivity of $LiTMO_2$ due to the softening of interlayer bonding as well as the increase of phonon-vacancy scattering; (3) the thermal conductivity of LiTMO$_2$ polycrystals can be tuned by several times by tuning the grain size; (4) the internal compressive or tensile strain induced during charge/discharge cycles can enhance or reduce the LiTMO$_2$ thermal conductivity multiple times. These findings are expected to generate broad impact not only on the fundamental thermal physics of complex oxides but also on the development of next-generation batteries and memristors.

## Acknowledgements

This work is supported in part by Department of Energy grant DE-FG0209ER46554 and by the McMinn Endowment. Computations were performed at the National Energy Research Scientific Computing Center (NERSC), a Department of Energy, Office of Science, User Facility funded through Contract No. DE-AC02-05CH11231. Computations also used the Extreme Science and Engineering Discovery Environment (XSEDE).

## Author contributions

T.F. and S.T.P. conceived the project. T.F. conducted the simulations and wrote the manuscript. A.O. contributed to simulations and discussions. All authors contributed to revising the manuscript.

## Additional information

Supplementary information including computational methodology is available in the online version of the paper.

Correspondence and requests for materials should be addressed to T.F. and S.T.P.

## Competing financial interests

The authors declare no competing financial interests.

# Supplementary Information

## Quantum Prediction of Ultra-Low Thermal Conductivity in Lithium Intercalation Materials

Tianli Feng[1,2,*,a], Andrew O'Hara[1], Sokrates T. Pantelides[1,2,†]

*[1]Department of Physics and Astronomy and Department of Electrical Engineering and Computer Science, Vanderbilt University, Nashville, Tennessee 37235, USA*

*[2]Center for Nanophase Materials Sciences, Oak Ridge National Laboratory, Oak Ridge, Tennessee 37831, USA*

*email: fengt@ornl.gov

†email: pantelides@vanderbilt.edu

[a] Present address: Energy and Transportation Science Division, Oak Ridge National Laboratory, Oak Ridge, Tennessee 37831, USA

**Table of Contents**

# Section 1. Methods

## A. Three- and four-phonon scattering rates calculations

The three- and four-phonon scattering rates, $\tau_{3,\lambda}^{-1}$ and $\tau_{4,\lambda}^{-1}$ are calculated by Fermi's golden rule (FGR)[1]:

$$\tau_{3,\lambda}^{-1} = \sum_{\lambda'\lambda''} \left[ \frac{1}{2}(1 + n_{\lambda'}^0 + n_{\lambda''}^0)\mathcal{L}_- + (n_{\lambda'}^0 - n_{\lambda''}^0)\mathcal{L}_+ \right], \tag{S.1}$$

$$\tau_{4,\lambda}^{-1} = \sum_{\lambda'\lambda''\lambda'''} \left[ \frac{1}{6} \frac{n_{\lambda'}^0 n_{\lambda''}^0 n_{\lambda'''}^0}{n_\lambda^0} \mathcal{L}_{--} + \frac{1}{2} \frac{(1+n_{\lambda'}^0)n_{\lambda''}^0 n_{\lambda'''}^0}{n_\lambda^0} \mathcal{L}_{+-} + \frac{1}{2} \frac{(1+n_{\lambda'}^0)(1+n_{\lambda''}^0)n_{\lambda'''}^0}{n_\lambda^0} \mathcal{L}_{++} \right], \tag{S.2}$$

Here $\lambda$ stands for $(\mathbf{q}, j)$ with $\mathbf{q}$ and $j$ labeling the phonon wave vector and dispersion branch, respectively. $n^0 = \left(e^{\frac{\hbar\omega}{k_B T}} - 1\right)^{-1}$ is the phonon occupation number, and $\omega$ is the phonon angular frequency. The transition probability matrix $\mathcal{L}$ is determined by the third-order and fourth-order interatomic force constants (IFCs) [1]:

$$\mathcal{L}_\pm = \frac{\pi\hbar}{4N_\mathbf{q}} \left|V_\pm^{(3)}\right|^2 \Delta_\pm \frac{\delta(\omega_\lambda \pm \omega_{\lambda'} - \omega_{\lambda''})}{\omega_\lambda \omega_{\lambda'} \omega_{\lambda''}}, \tag{S.3}$$

$$\mathcal{L}_{\pm\pm} = \frac{\pi\hbar}{4N_\mathbf{q}} \frac{\hbar}{2N_\mathbf{q}} \left|V_{\pm\pm}^{(4)}\right|^2 \Delta_{\pm\pm} \frac{\delta(\omega_\lambda \pm \omega_{\lambda'} \pm \omega_{\lambda''} - \omega_{\lambda'''})}{\omega_\lambda \omega_{\lambda'} \omega_{\lambda''} \omega_{\lambda'''}}, \tag{S.4}$$

where $V_\pm^{(3)}$ and $V_{\pm\pm}^{(4)}$ are

$$V_\pm^{(3)} = \sum_{b,l_1 b_1, l_2 b_2} \sum_{\alpha\alpha_1\alpha_2} \phi_{0b,l_1 b_1,l_2 b_2}^{\alpha\alpha_1\alpha_2} \frac{e_{\alpha b}^\lambda e_{\alpha_1 b_1}^{\pm\lambda'} e_{\alpha_2 b_2}^{-\lambda''}}{\sqrt{\bar{m}_b \bar{m}_{b_1} \bar{m}_{b_2}}} e^{(\pm i\mathbf{q}'\cdot\mathbf{r}_{l_1} - i\mathbf{q}''\cdot\mathbf{r}_{l_2})}, \tag{S.5}$$

$$V_{\pm\pm}^{(4)} = \sum_{b,l_1 b_1, l_2 b_2, l_3 b_3} \sum_{\alpha\alpha_1\alpha_2\alpha_3} \phi_{0b,l_1 b_1,l_2 b_2,l_3 b_3}^{\alpha\alpha_1\alpha_2\alpha_3} \frac{e_{\alpha b}^\lambda e_{\alpha_1 b_1}^{\pm\lambda'} e_{\alpha_2 b_2}^{\pm\lambda''} e_{\alpha_3 b_3}^{-\lambda'''}}{\sqrt{\bar{m}_b \bar{m}_{b_1} \bar{m}_{b_2} \bar{m}_{b_3}}} e^{(\pm i\mathbf{q}'\cdot\mathbf{r}_{l_1} \pm i\mathbf{q}''\cdot\mathbf{r}_{l_2} - i\mathbf{q}'''\cdot\mathbf{r}_{l_3})}, \tag{S.6}$$

$N_\mathbf{q}$ is the total number of $\mathbf{q}$ points. The Kronecker deltas $\Delta_\pm = \Delta_{\mathbf{q}\pm\mathbf{q}'-\mathbf{q}'',\mathbf{R}}$ and $\Delta_{\pm\pm} = \Delta_{\mathbf{q}\pm\mathbf{q}'\pm\mathbf{q}''-\mathbf{q}''',\mathbf{R}}$ describe the momentum selection rule and have the property that $\Delta_{m,n} = 1$ (if $m = n$), or 0 (if $m \neq n$). $\mathbf{R}$ is a reciprocal lattice vector. $\phi^{\alpha\alpha_1\alpha_2}_{0b,l_1b_1,l_2b_2}$ and $\phi^{\alpha\alpha_1\alpha_2\alpha_3}_{0b,l_1b_1,l_2b_2,l_3b_3}$ are the third- and fourth-order force constants, which are calculated from DFT. $l$, $b$, and $\alpha$ label the indices of unit cells, basis atoms, and Cartesian directions, respectively. $\mathbf{r}_l$ is the position of the unit cell $l$. $\bar{m}_b$ is the average atomic mass at the lattice site $b$. $e$ is the phonon eigenvector component. The eigenvectors are solved by using the dynamical matrix shown in Phonopy[2]. Details of correctly using the eigenvectors together with the phase term $\exp(i\mathbf{q}\cdot\mathbf{r})$ can be found in Ref.[3].

The delta function $\delta(\Delta\omega)$ in the calculation of each $\mathcal{L}$ is evaluated by using the adaptive Gaussian broadening method, which has been established for three-phonon scattering in Ref.[4]. Similar to three-phonon processes, in this work, we establish the adaptive Gaussian broadening method for four-phonon processes as

$$\delta(\omega_\lambda \pm \omega_{\lambda'} \pm \omega_{\lambda''} - \omega_{\lambda'''}) = \frac{1}{\sqrt{\pi}\sigma} \exp\left(-\frac{(\omega_\lambda \pm \omega_{\lambda'} \pm \omega_{\lambda''} - \omega_{\lambda'''})^2}{\sigma^2}\right), \quad \text{(S.7)}$$

$$\sigma \sim |\pm\mathbf{v}_{\lambda'} \pm \mathbf{v}_{\lambda''} - \mathbf{v}_{\lambda'''}||\Delta q|, \quad \text{(S.8)}$$

$$|\mathbf{v}||\Delta q| = \sqrt{\frac{|\mathbf{v}\cdot\Delta\mathbf{q}_1|^2 + |\mathbf{v}\cdot\Delta\mathbf{q}_2|^2 + |\mathbf{v}\cdot\Delta\mathbf{q}_3|^2}{6}}, \quad \text{(S.9)}$$

where $\Delta\mathbf{q}_1, \Delta\mathbf{q}_2$, and $\Delta\mathbf{q}_3$ are the three vectors of a unit grid of the $\mathbf{q}$ mesh.

## B. Lattice thermal conductivity calculations

The thermal conductivity along a crystal orientation $\kappa_\alpha$ ($\alpha = x, y, z$) is calculated by using the exact solution to the Boltzmann transport equation[5,6],

$$\kappa_\alpha = \frac{1}{V} \sum_\lambda v_{\alpha,\lambda}^2 c_\lambda \tau_\lambda^{it}, \tag{S.10}$$

where $V$ is crystal volume, $v_\alpha$ is phonon group velocity projection along the direction $\alpha$, and $c_\lambda$ is phonon specific heat per mode. $\tau_\lambda^{it}$ is the phonon relaxation time solved by an iterative scheme:

$$\tau_\lambda^{it} = \tau_\lambda (1 + \Theta_\lambda), \tag{S.11}$$

$$\frac{1}{\tau_\lambda} = \tau_{3,\lambda}^{-1} + \tau_{4,\lambda}^{-1} + \sum_{\lambda' \neq \lambda} \Gamma_{\lambda\lambda'}^{iso}, \tag{S.12}$$

$$\Gamma_{\lambda\lambda'}^{iso} = \frac{\pi}{2N_c} \omega_\lambda \omega_{\lambda'} \sum_b^n g_b |\mathbf{e}_\lambda^b \cdot \mathbf{e}_{\lambda'}^{b*}|^2 \delta(\omega_\lambda - \omega_{\lambda'}), \tag{S.13}$$

$$g_b = \sum_i f_{ib} \left(1 - \frac{m_{ib}}{\bar{m}_b}\right)^2, \tag{S.14}$$

$$\Theta_\lambda = \sum_{\lambda'\lambda''}^{(+)} \mathcal{L}_+ \left(\xi_{\lambda\lambda''} \tau_{\lambda''}^{it} - \xi_{\lambda\lambda'} \tau_{\lambda'}^{it}\right) + \sum_{\lambda'\lambda''}^{(-)} \frac{1}{2} \mathcal{L}_- \left(\xi_{\lambda\lambda''} \tau_{\lambda''}^{it} + \xi_{\lambda\lambda'} \tau_{\lambda'}^{it}\right) + \sum_{\lambda' \neq \lambda} \Gamma_{\lambda\lambda'}^{iso} \xi_{\lambda\lambda'} \tau_{\lambda'}^{it}, \tag{S.15}$$

where $\xi_{\lambda\lambda'} = v_{\lambda',\alpha} \omega_{\lambda'} / v_{\lambda,\alpha} \omega_\lambda$. Equation (S.11) is solved iteratively because both the left and the right-hand sides contain the unknown variable $\tau_\lambda^{it}$, and thus the method is called Iterative Scheme. $g_b$ characterizes the magnitude of mass disorder induced by isotopes or Li random distribution in lattice, where $i$ indicates isotope types, $f_{ib}$ is the fraction of isotope $i$ in lattice sites of basis atom $b$, $m_{ib}$ is the mass of isotope $i$, $\bar{m}_b$ is the average atom mass of basis $b$ sites. $\mathbf{e}$ is the phonon eigenvectors. The thermal conductivities w/o four-phonon scattering is determined by including the $\tau_{4,\lambda}^{-1}$ term in Eq. (S.12) or not.

## C. Phonon scattering induced by lithium disorder

For delithiated $Li_xTMO_2$ (x<1), the lithium atoms are randomly distributed. The random distribution of Li atoms induces an extra scattering term, phonon-Li disorder scattering, in addition to the intrinsic phonon-phonon scattering. This disorder is to redistribute Li from ordered patterns to random patterns as shown in Supplementary Fig. S2. Since the concentration of Li does not change with disorder, this redistribution does not affect the interlayer bonding strength or anharmonicity of $Li_xTMO_2$. The only effect of this Li redistribution is to contribute a mass-disorder scattering term as described by Eq. (S.13). This effect is small for the thermal transport in $Li_xTMO_2$ since the vibration of Li does not contribute much to the thermal conductivity. The thermal conductivity of $Li_xTMO_2$ is dominated by the frequency range 0~5 THz, while the Li vibration frequency is above this range (the phonon DOS of Li is nearly 0 in this frequency range).

## D. Density functional theory simulation details

DFT calculations are performed using the Vienna Ab initio simulation package (VASP)[7] using the local density approximation (LDA) for exchange and correlation and the projector-augmented-wave method[8]. It has been shown that LDA and generalized gradient approximation with the Hubbard model correction (GGA+U) yield similar phonon dispersions without soft modes and analogous thermodynamic functions for $LiTMO_2$.[9] The plane-wave energy cutoff is 500 eV. The energy convergence threshold is set at $10^{-8}$ eV. In the cell relaxation, the force convergence threshold is $10^{-7}$ eV/Å.

The phonon dispersion relations are calculated using Phonopy[2] with a finite difference method. The supercell sizes for the calculations of second-order force constants are 3×3×3 (108 atoms), 3×3×3 (297 atoms), 3×4×3 (252 atoms), 3×3×3 (270 atoms), 4×4×2 (256 atoms), and 4×2×2 (224 atoms) for $LiCoO_2$, $Li_{0.67}CoO_2$, $Li_{0.5}CoO_2$, $Li_{0.33}CoO_2$, $LiNbO_2$, and $Li_{0.5}NbO_2$, respectively.

The supercell sizes for the calculations of third-order force constants are 3×3×3 (108 atoms), 2×2×2 (88 atoms), 2×3×2 (84 atoms), 2×2×2 (80 atoms), 3×3×2 (144 atoms), and 4×2×1 (112 atoms) for $LiCoO_2$, $Li_{0.67}CoO_2$, $Li_{0.5}CoO_2$, $Li_{0.33}CoO_2$, $LiNbO_2$, and $Li_{0.5}NbO_2$, respectively. The supercell sizes for the calculations of fourth-order force constants for $LiCoO_2$, $Li_{0.33}CoO_2$, $LiNbO_2$, and $Li_{0.5}NbO_2$, are the same as those used in third-order force constants calculations. The cutoff radii are chosen as 6th and 2nd nearest neighbors for the third and fourth order force constants calculations, respectively.

## Section 2. Counts of force constants

An $n^{th}$ order force constant can be written as $\Phi_{0,b_1;\ l_2,b_2;\ ...;\ l_n,b_n}^{\alpha_1;\alpha_2;...;\alpha_n}$, which describe the anharmonic force constant among the atoms $(0, b_1), (l_2, b_2), ..., (l_n, b_n)$ in the $(\alpha_1; \alpha_2; ...; \alpha_n)$ direction. $(l_n, b_n)$ represents the $b_n$-th atom in the $l_n$-th unit cell. It is noted that in all the force constants, $\Phi_{0,b_1;\ l_2,b_2;\ ...;\ l_n,b_n}^{\alpha_1;\alpha_2;...;\alpha_n}$, the first atom must be chosen in the origin unit cell, "0", (the remaining $n-1$ atoms can go over all unit cells). If at least one of the $n-1$ atoms is far away from the origin atom $(0, b_1)$, the force constant value $\Phi_{0,b_1;\ l_2,b_2;\ ...;\ l_n,b_n}^{\alpha_1;\alpha_2;...;\alpha_n}$ will be trivial (~ 0). Usually, this threshold of distance is called cutoff radius, beyond which the force constant values are trivial and negligible. For the third-order force constants, the cutoff radius is usually to the $5^{th}$-$7^{th}$ nearest neighbors. For the fourth-order force constants, it can be much shorter, i.e., the $2^{nd}$ nearest neighbors[10]. In the present study, all the cutoff radii we choose exceed these ranges and thus cover all significant force constant values.

Therefore, the counts of significant (values > 1 eV/Å) force constant in Fig. 2 do not depend on the size of the supercell one chooses, and they can be interpreted quantitatively. The counts of trivial force constants (values ~ 0), the bottom bars in each subfigure in Fig. 2, should not be interpreted quantitatively since these counts contain the force constants among the atoms that are far away from each other, and therefore depend arbitrarily on the sizes of supercells one chooses.

## Section 3. Thermal conductivity orientational averaging for polycrystals

Single-crystal grains stack randomly together to form polycrystals, whose thermal conductivity is a random-orientation average of the anisotropic thermal conductivities of single crystal grains. Two rough models are often used to do the average, i.e., conductive averaging

$$\kappa_a^{para} = \frac{2}{3}\kappa_\parallel + \frac{1}{3}\kappa_\perp \tag{S.16}$$

and resistive averaging

$$\kappa_a^{seri} = \left(\frac{2}{3}\kappa_\parallel^{-1} + \frac{1}{3}\kappa_\perp^{-1}\right)^{-1} \tag{S.17}$$

where $\kappa_\parallel$ and $\kappa_\perp$ are the in-plane and through-plane thermal conductivities of the single crystal. The former assumes the grains are paved in parallel to each other, and the heat flows through all grains independently in parallel. Therefore, it can estimate the through-plane thermal conductivity of a thin film. The latter assumes the grains are paved in serial, and the heat flows through all grains in serial. This can estimate the along-wire thermal conductivity of a thin wire.

However practical polycrystals are neither thin films or thin wires, and the thermal conductivity is located in between $\kappa_a^{para}$ and $\kappa_a^{seri}$. Ref.[11] has done a comprehensive study and revealed that the thermal conductivity of polycrystals can be well estimated by

$$\kappa_a = \frac{2}{3}\kappa_\parallel + \frac{1}{3}\kappa_\perp - \frac{2}{9}\frac{(r-1)^2}{r+2}\kappa_\parallel, \tag{S.18}$$

which, therefore, is adopted in this work for the calculations of averaged thermal conductivity for polycrystals. In addition to the averaging, phonon grain-boundary is also taken into account. The grain boundary scattering is calculated by $\tau_b^{-1} = v_\lambda/D$, where $D$ is the average grain diameter.

# Section 4. Atomic structures of Li$_x$CoO$_2$

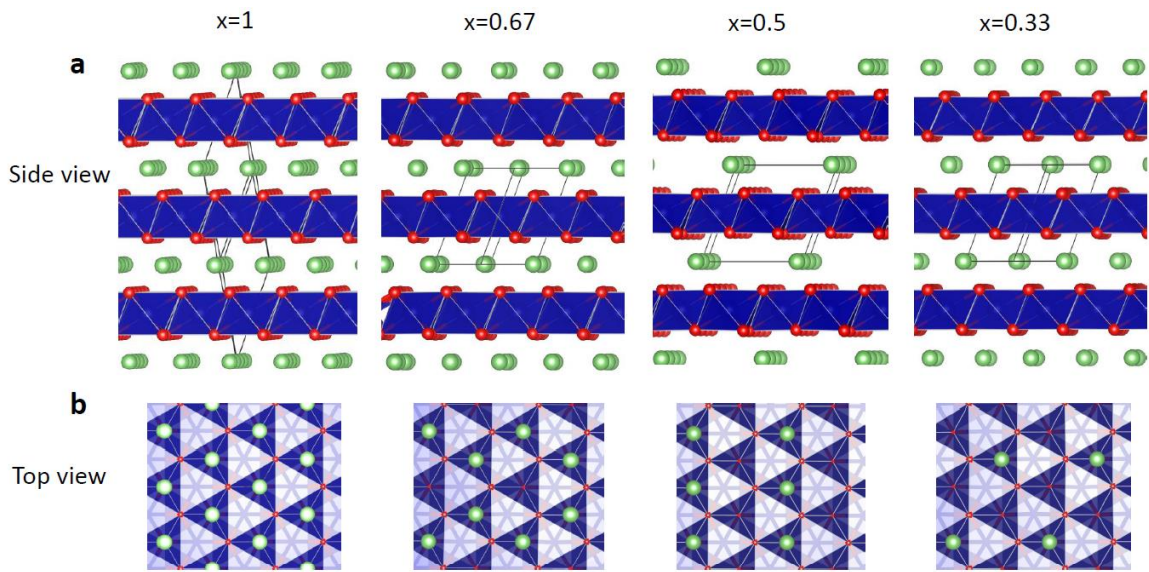

Fig. S1. Single-crystal structures of Li$_x$CoO$_2$ used in this work. **a**, Side view. **b**, Top view.

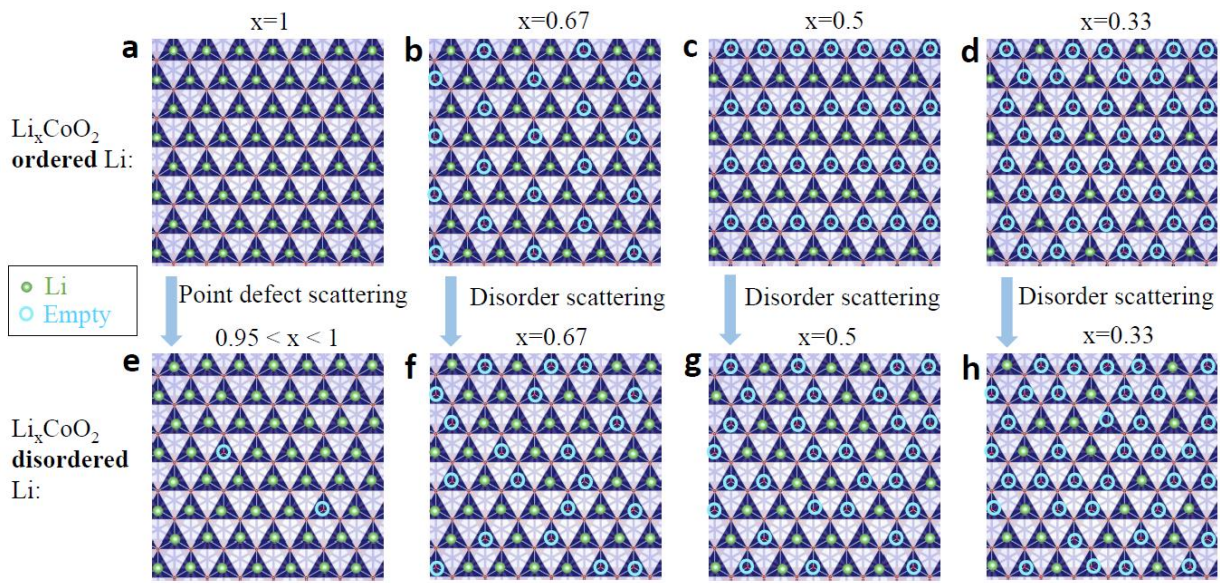

Fig. S2. Top view of Li$_x$CoO$_2$ crystals with ordered Li (**a-d**) and disordered Li (**e-h**).

# Section 5. Phonon density of states of Li$_x$CoO$_2$

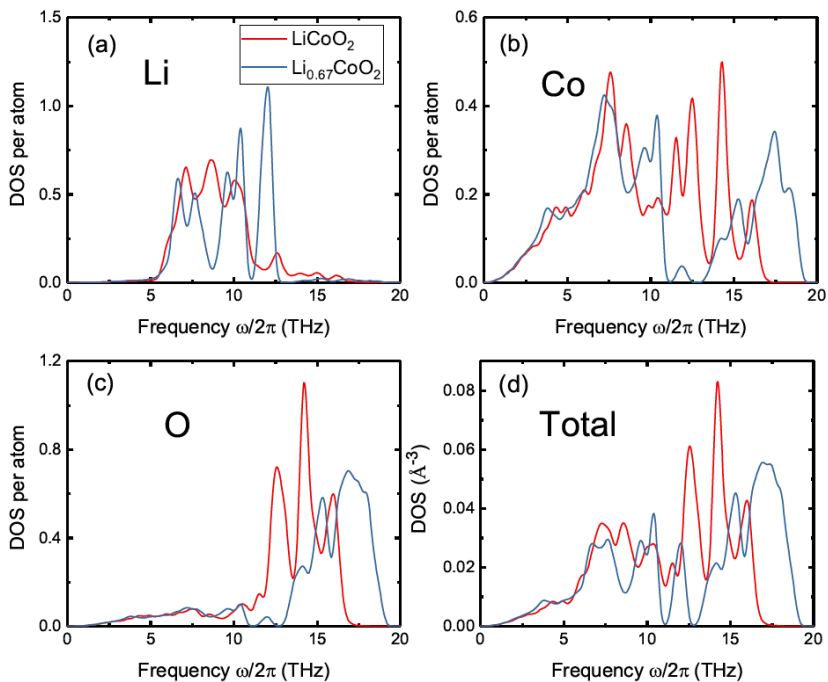

Fig. S3. Comparison of phonon densities of states of LiCoO$_2$ and Li$_{0.67}$CoO$_2$.

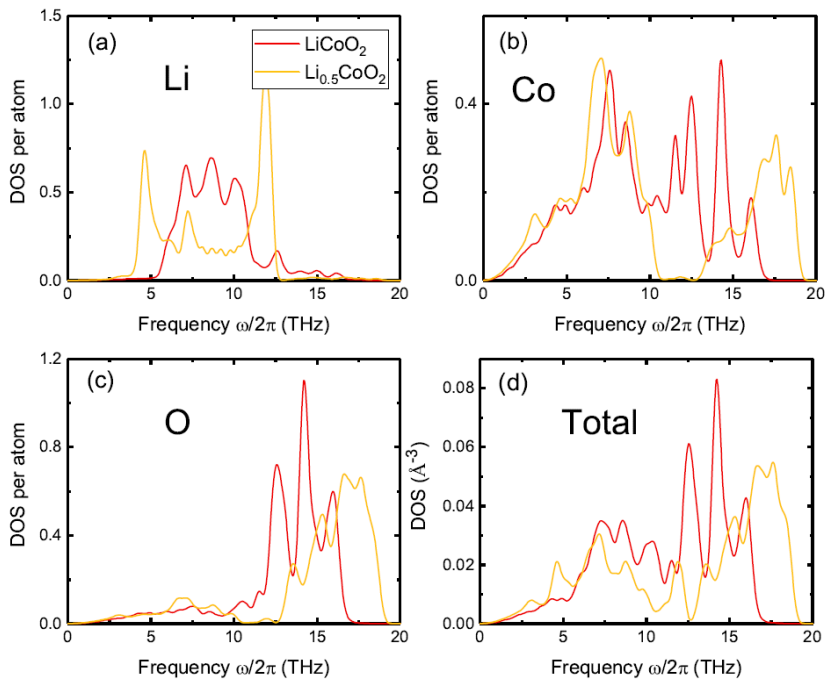

Fig. S4. Comparison of phonon densities of states of LiCoO$_2$ and Li$_{0.5}$CoO$_2$.

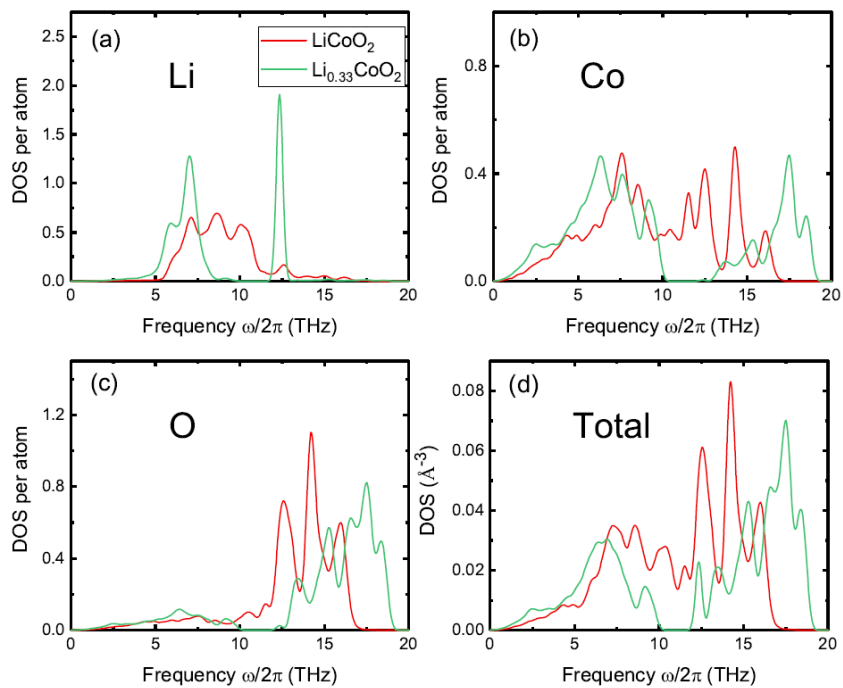

Fig. S5. Comparison of phonon densities of states of LiCoO$_2$ and Li$_{0.33}$CoO$_2$.

## Section 6. Phonon group velocities of Li$_x$CoO$_2$

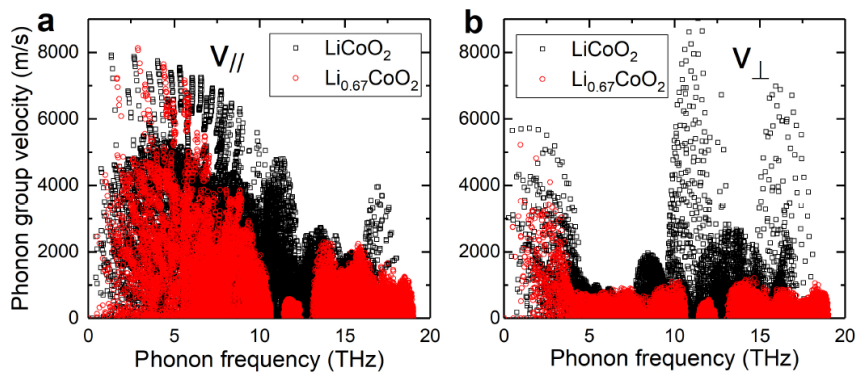

Fig. S6. Comparison of in-plane (**a**) and through-plane (**b**) phonon group velocities of LiCoO$_2$ and Li$_{0.67}$CoO$_2$.

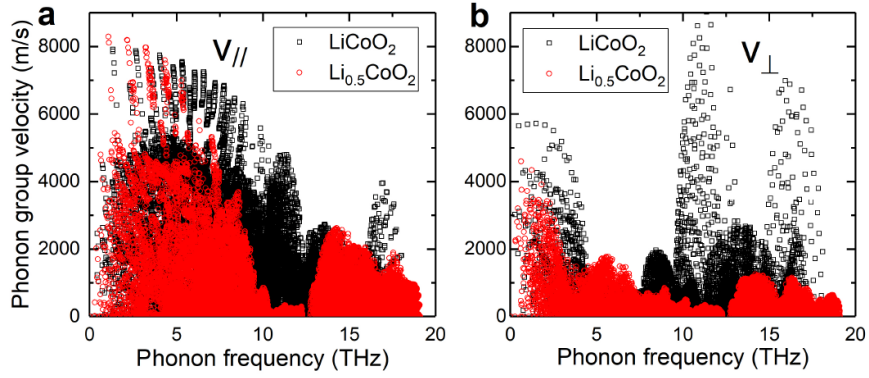

Fig. S7. Comparison of in-plane (**a**) and through-plane (**b**) phonon group velocities of $LiCoO_2$ and $Li_{0.5}CoO_2$.

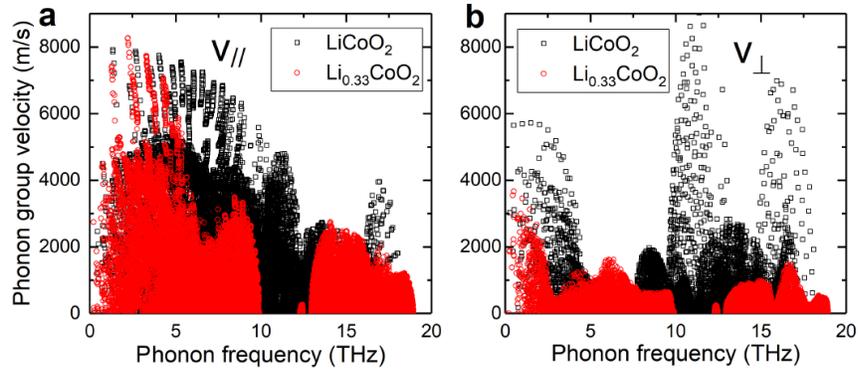

Fig. S8. Comparison of in-plane (**a**) and through-plane (**b**) phonon group velocities of $LiCoO_2$ and $Li_{0.33}CoO_2$.

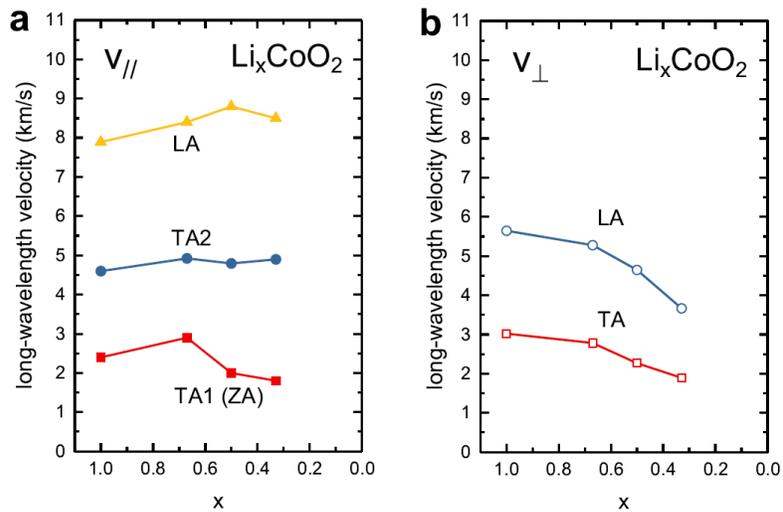

Fig. S9. In-plane (**a**) and through-plane (**b**) long-wavelength acoustic phonon velocities of $Li_xCoO_2$ as a function of x.

## Section 7. Specific heat of $Li_xCoO_2$

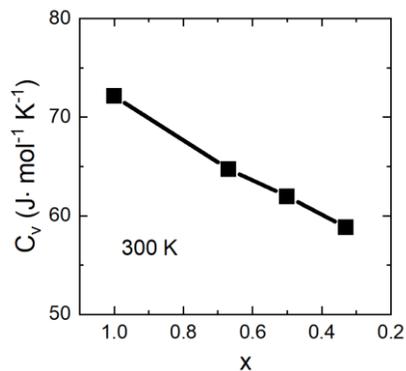

Fig. S10. Specific heat of $Li_xCoO_2$ as a function of Li concentration.

## Section 8. Atomic structures of $Li_xNbO_2$

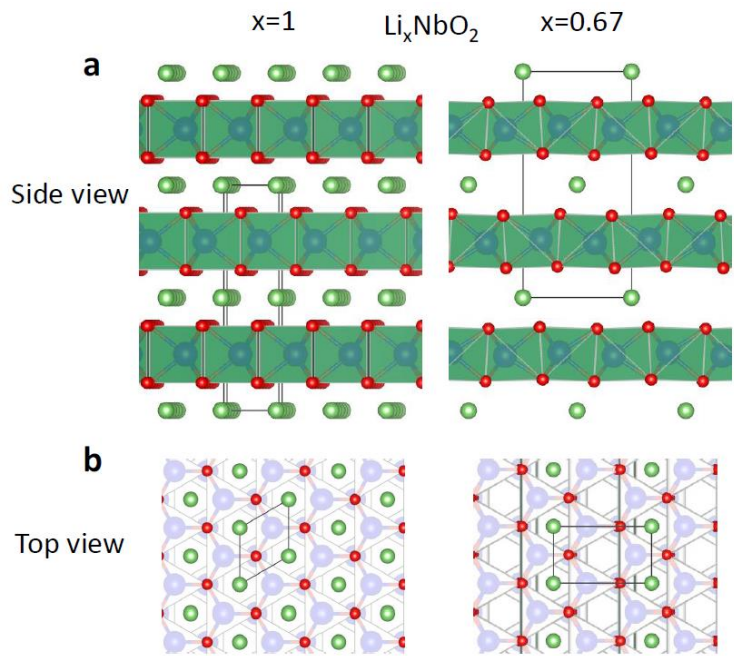

Fig. S11. Single-crystal structures of $Li_xNbO_2$ used in this work. **a**, Side view. **b**, Top view.

## Section 9. Phonon dispersions of Li$_x$NbO$_2$

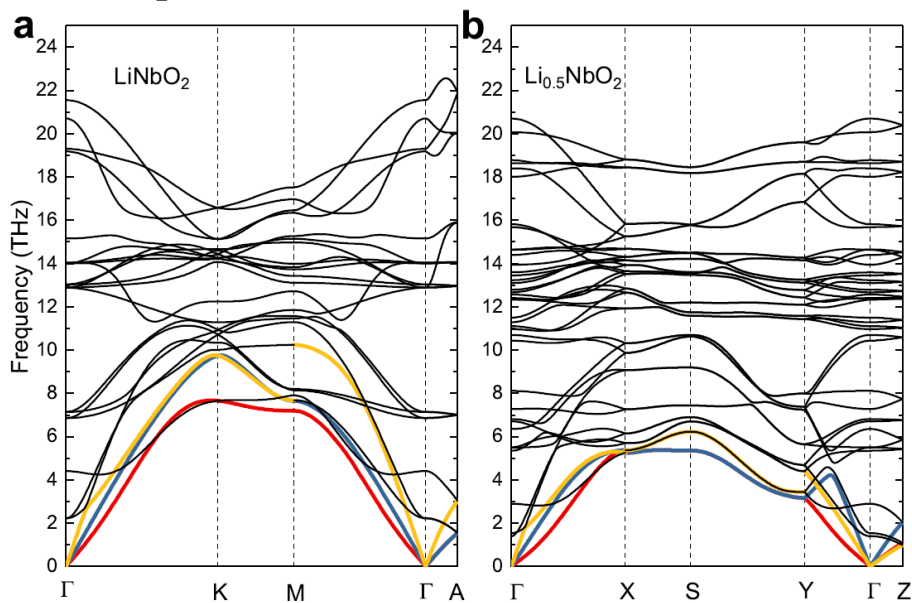

Fig. S12. Comparison of phonon dispersion relations of LiNbO$_2$ and Li$_{0.5}$NbO$_2$. Colored thick curves highlight acoustic branches.

## Section 10. Density of states of Li$_x$NbO$_2$

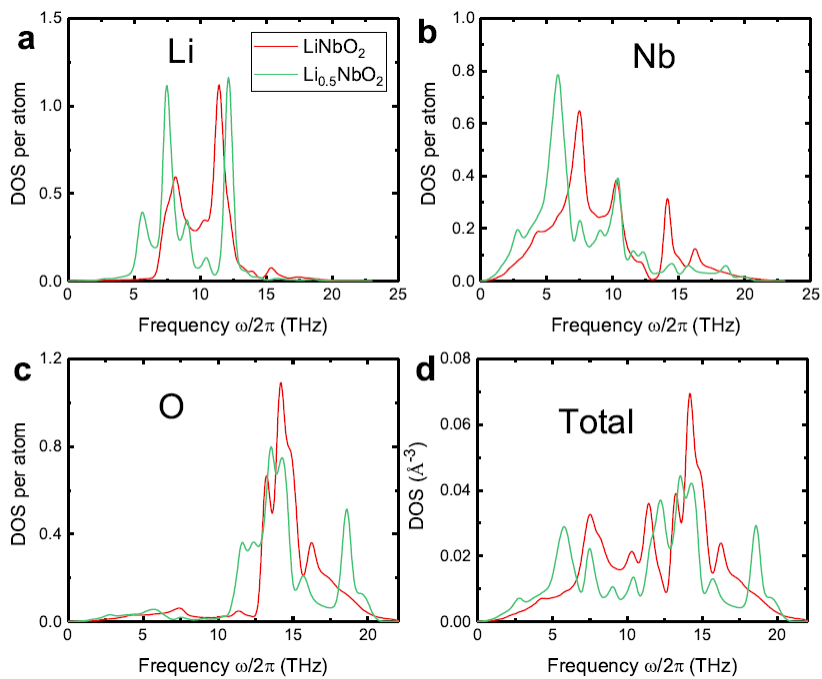

Fig. S13. Comparison of densities of states of LiNbO$_2$ and Li$_{0.5}$NbO$_2$.

## Section 11. Specific heat of Li$_x$NbO$_2$

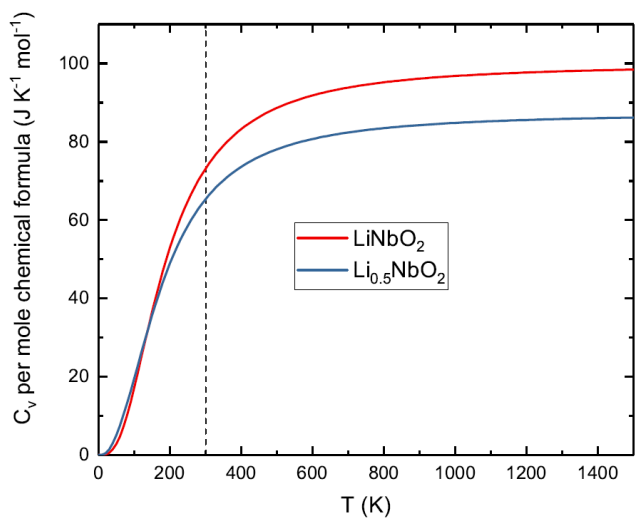

Fig. S14. Temperature-dependent lattice specific heats of LiNbO$_2$ and Li$_{0.5}$NbO$_2$.

## Section 12. Phonon group velocities of Li$_x$NbO$_2$

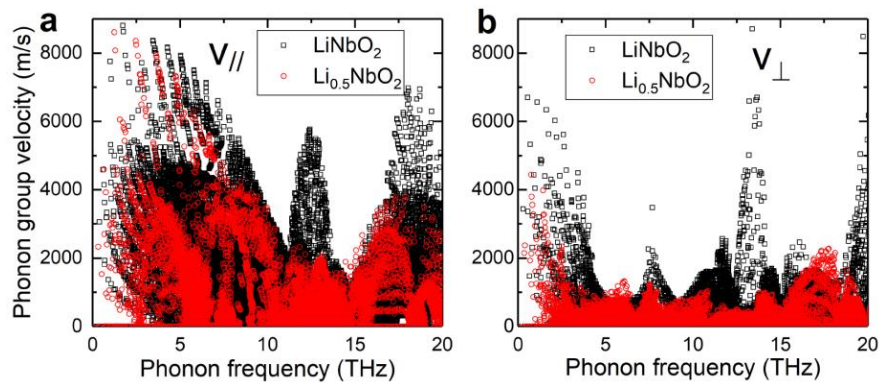

Fig. S15. Comparison of (**a**) in-plane and (**b**) through-plane phonon group velocities of LiNbO$_2$ and Li$_{0.5}$NbO$_2$.

# Section 13. Individual four-phonon scattering rates of $Li_xCoO_2$ and $Li_xNbO_2$

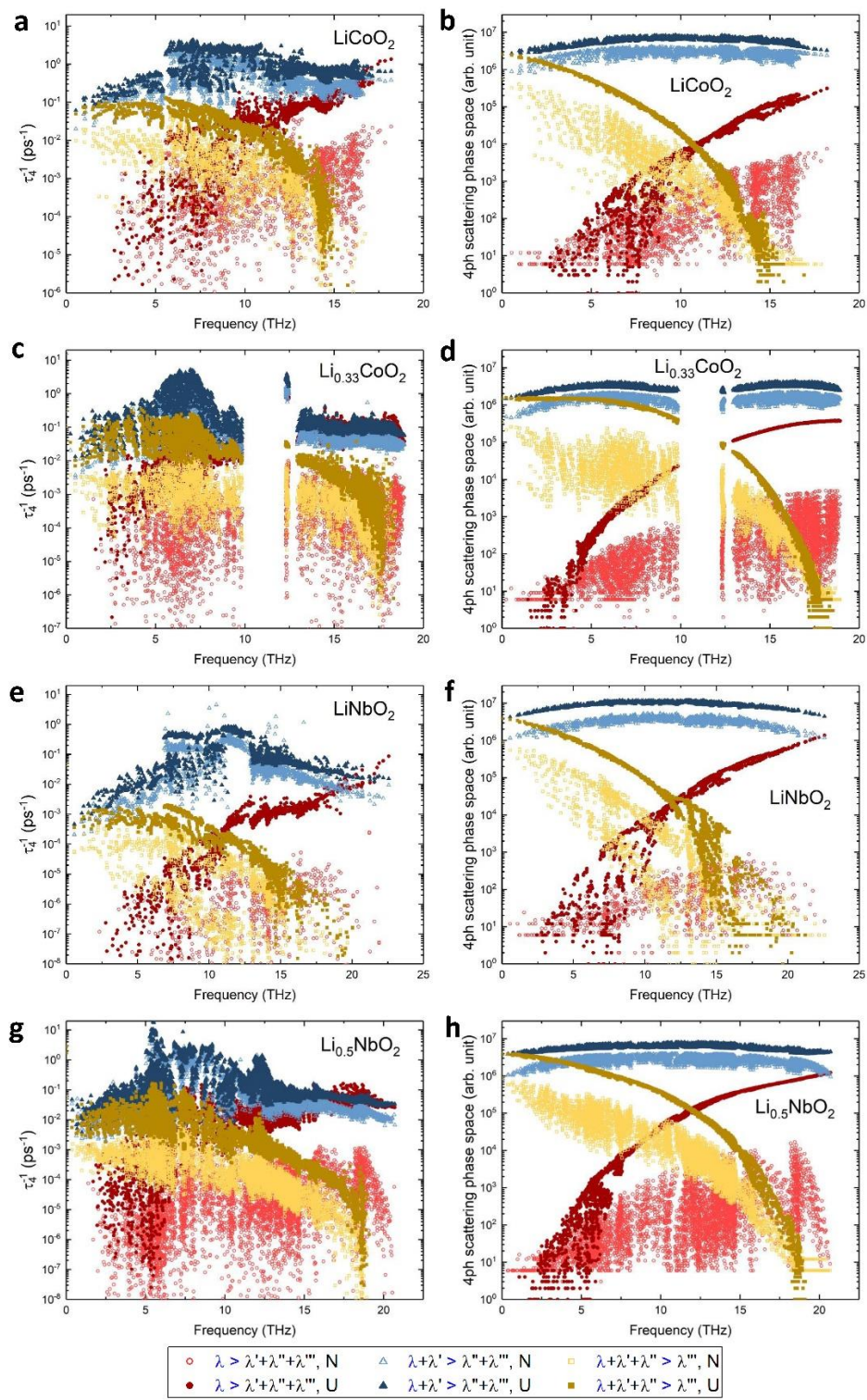

Fig. S16. Individual four-phonon scattering rates of $Li_xCoO_2$ and $Li_xNbO_2$.

## Section 14. Spectral thermal conductivities of $Li_xCoO_2$ and $Li_xNbO_2$

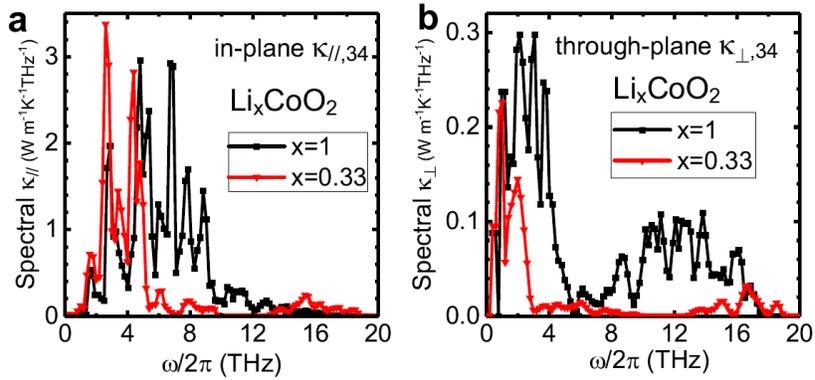

Fig. S17. Spectral thermal conductivity $\kappa(\omega)$ of $LiCoO_2$ and $Li_{0.33}CoO_2$ single crystals. **a**, In-plane thermal conductivity. **b**, Through-plane thermal conductivity. In $Li_{0.33}CoO_2$, Li disorder effect is included. The integration of the spectral $\kappa(\omega)$ gives the thermal conductivity $\kappa$, i.e., $\int_0^{\omega_{max}} \kappa(\omega)d\omega = \kappa$.

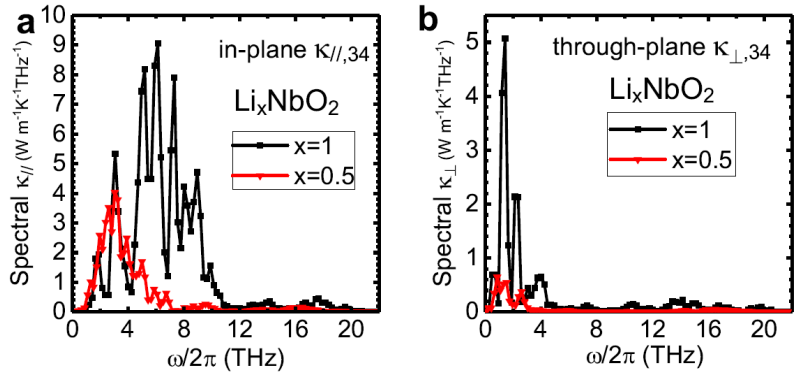

Fig. S18. Spectral thermal conductivity $\kappa(\omega)$ of $LiNbO_2$ and $Li_{0.5}NbO_2$ single crystals. **a**, In-plane thermal conductivity. **b**, Through-plane thermal conductivity. In $Li_{0.5}NbO_2$, Li disorder effect is included. The integration of the spectral $\kappa(\omega)$ gives the thermal conductivity $\kappa$, i.e., $\int_0^{\omega_{max}} \kappa(\omega)d\omega = \kappa$.

## Section 15. Phonon scattering rates and lifetimes of Li$_x$CoO$_2$

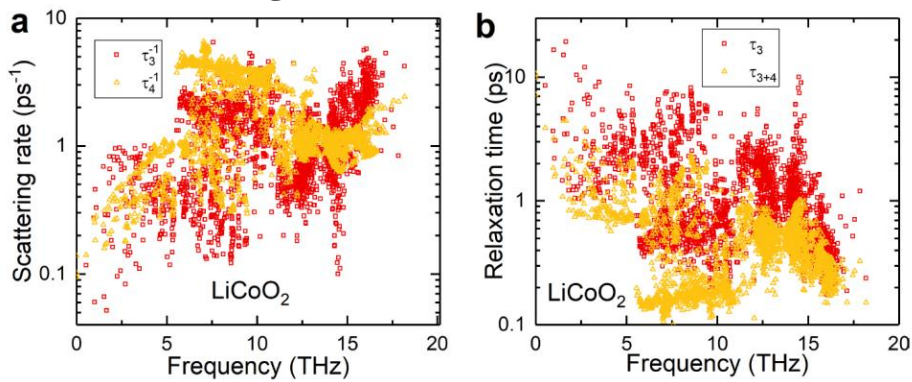

Fig. S19. Phonon scattering rates (**a**) and lifetimes (**b**) of LiCoO$_2$.

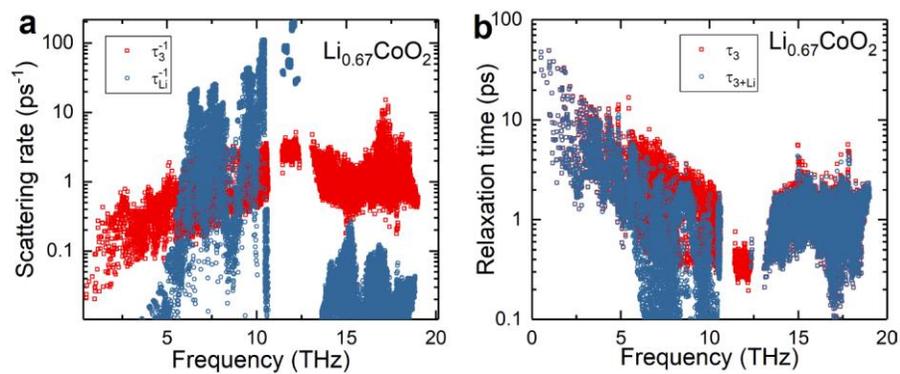

Fig. S20. Phonon scattering rates (**a**) and lifetimes (**b**) of Li$_{0.67}$CoO$_2$.

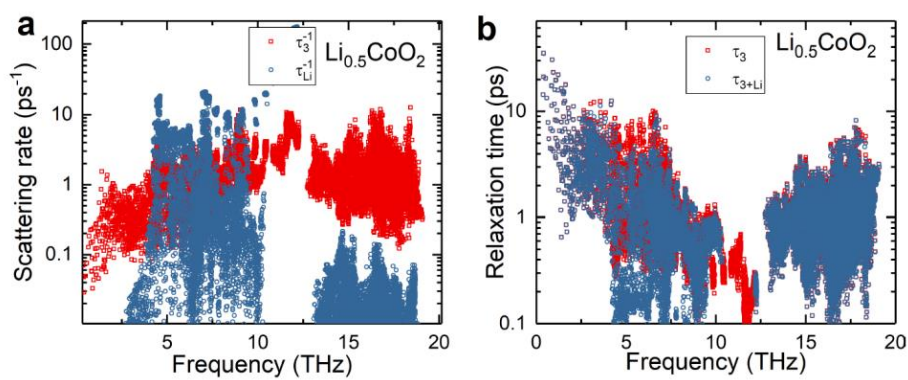

Fig. S21. Phonon scattering rates (**a**) and lifetimes (**b**) of Li$_{0.5}$CoO$_2$.

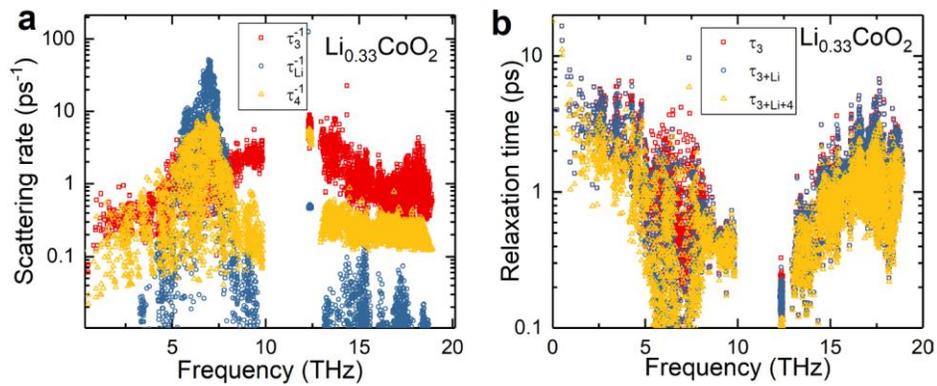

Fig. S22. Phonon scattering rates (**a**) and lifetimes (**b**) of $Li_{0.33}CoO_2$.

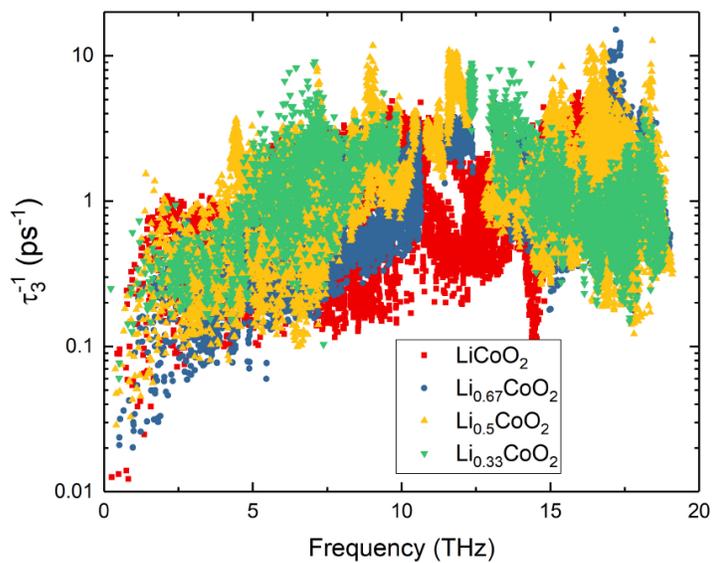

Fig. S23. Comparison of the three-phonon scattering rates of $Li_xCoO_2$ (x=1, 0.67, 0.5, 0.33).

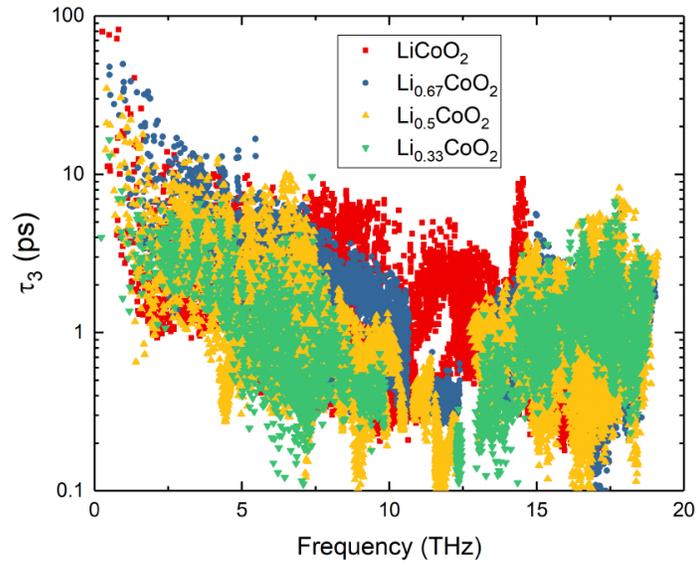

Fig. S24. Comparison of the three-phonon relaxation times of $Li_xCoO_2$ (x=1, 0.67, 0.5, 0.33).

## Section 16. Phonon scattering rates and lifetimes of $Li_xNbO_2$

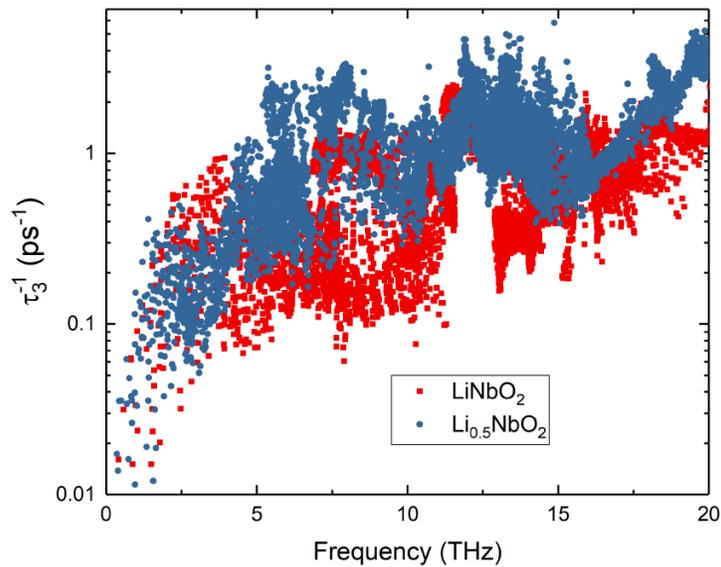

Fig. S26. Comparison of the three-phonon scattering rates of $Li_xNbO_2$ (x=1, 0.5).

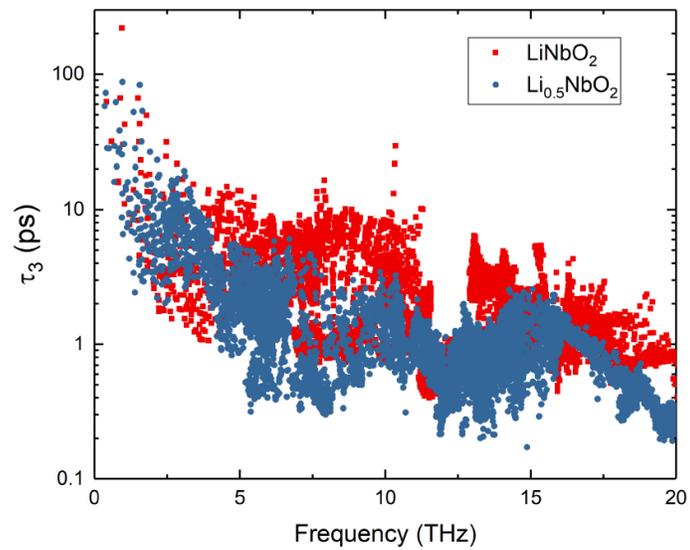

Fig. S27. Comparison of the three-phonon relaxation times of $Li_xNbO_2$ (x=1, 0.5).

# Section 17. Strain effect on Li$_{0.33}$CoO$_2$

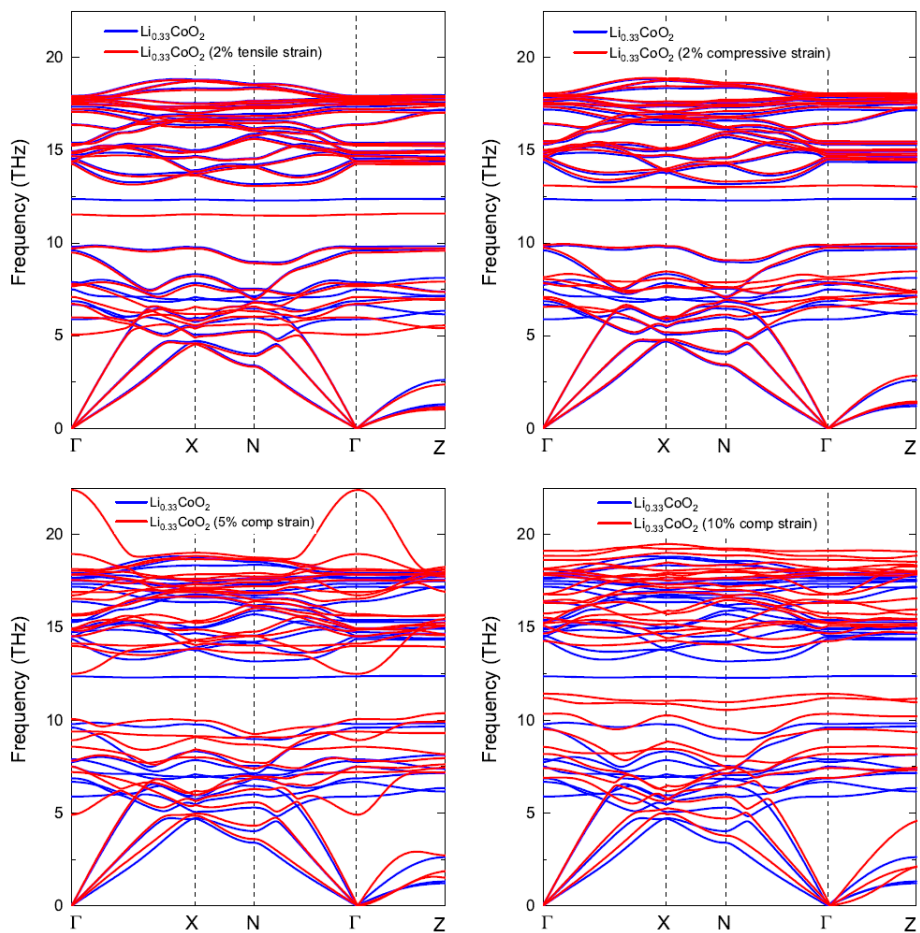

Fig. S28. Comparison of phonon dispersion relations of Li$_{0.33}$CoO$_2$ with (**a**) 2%, (**b**) -2%, (**c**) -5%, and (**d**) -10% through-plane strains.

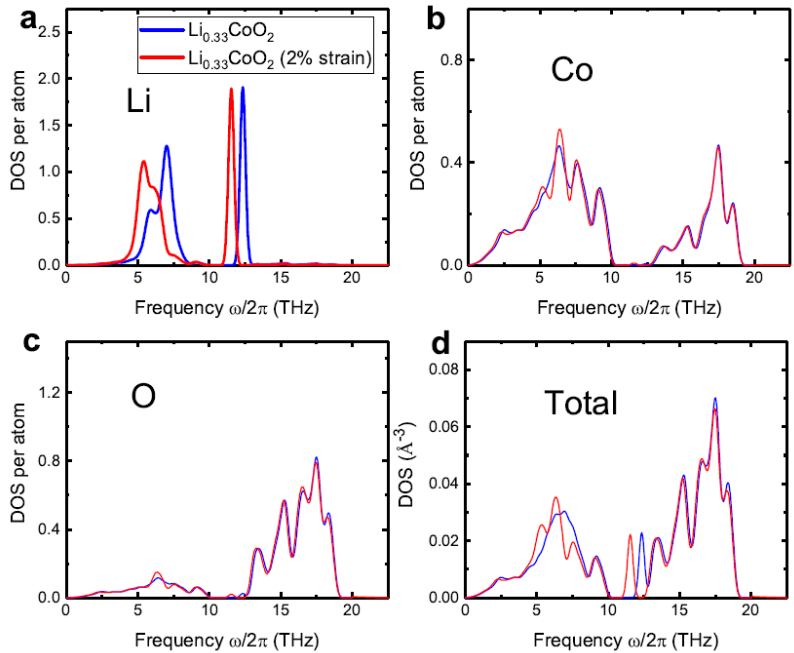

Fig. S29. Comparison of phonon density of states of $Li_{0.33}CoO_2$ w/o 2% through-plane strain.

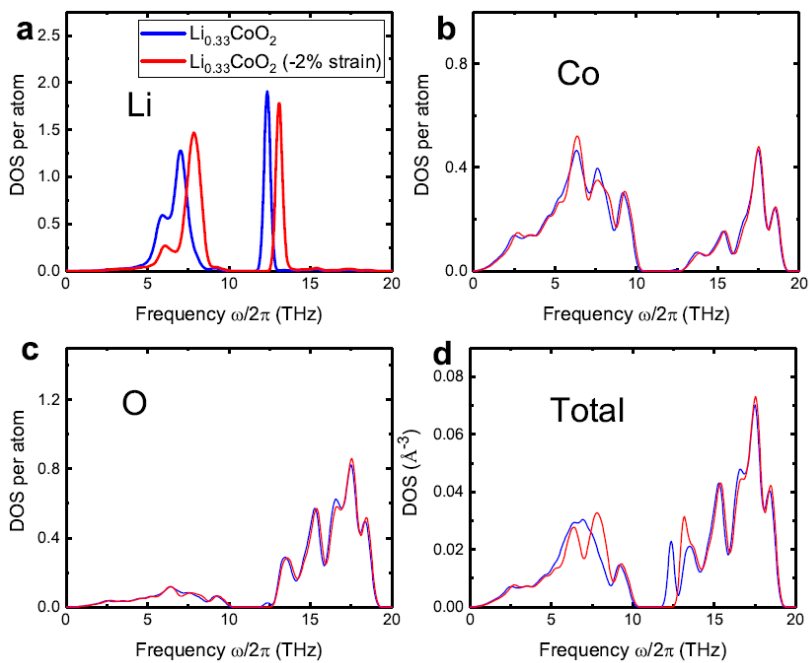

Fig. S30. Comparison of phonon density of states of $Li_{0.33}CoO_2$ w/o -2% through-plane strain.

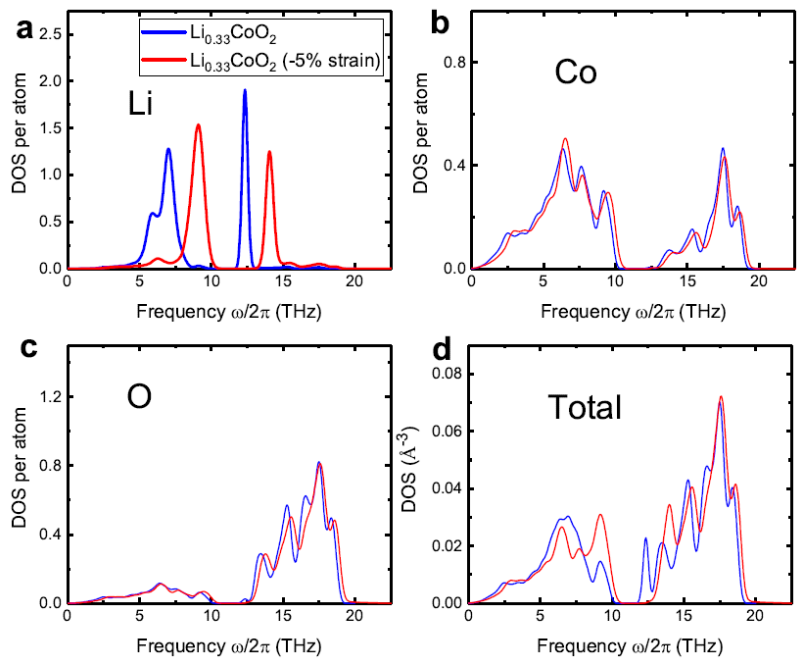

Fig. S31. Comparison of phonon density of states of $Li_{0.33}CoO_2$ w/o -5% through-plane strain.

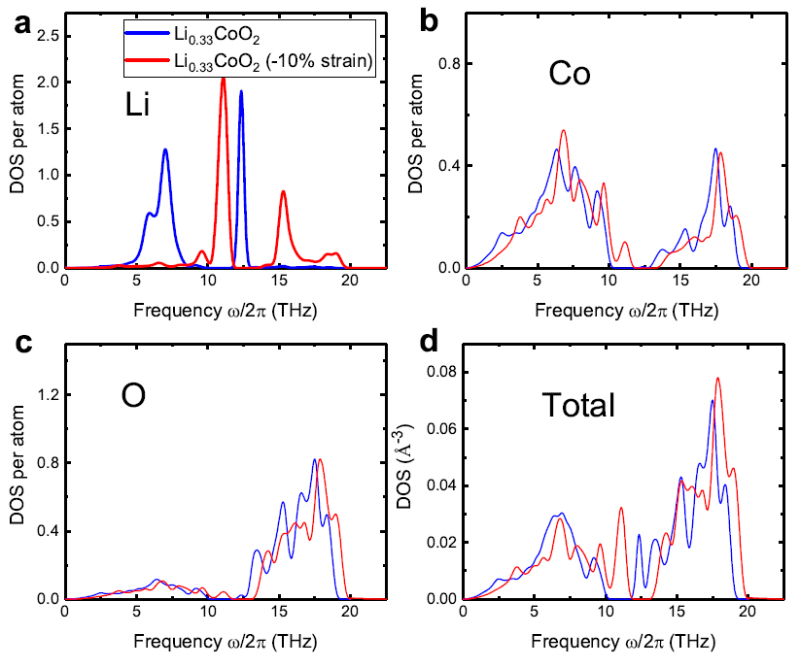

Fig. S32. Comparison of phonon density of states of $Li_{0.33}CoO_2$ w/o -10% through-plane strain.

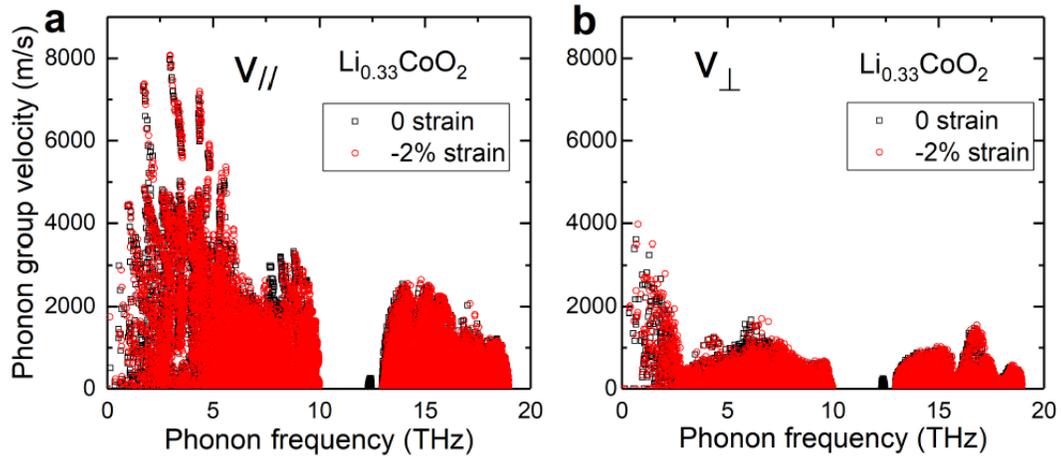

Fig. S33. Comparison of the (**a**) in-plane and (**b**) through-plane phonon group velocity of $Li_{0.33}CoO_2$ w/o

-2% through-plane strain.

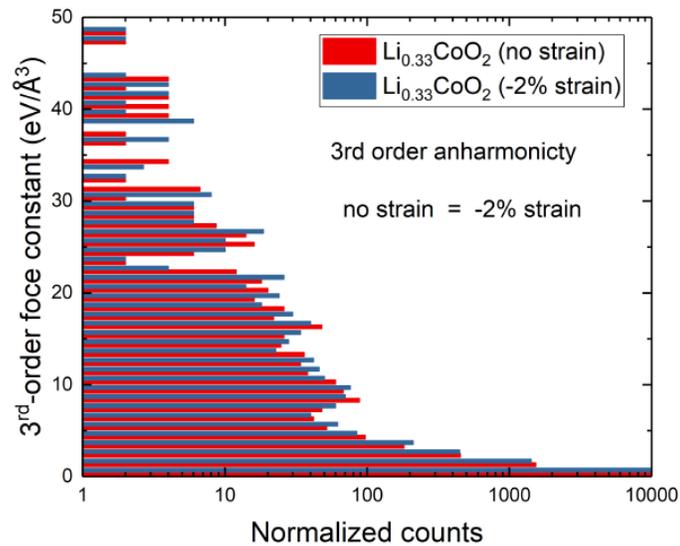

Fig. S34. Third-order force constants of $Li_{0.33}CoO_2$ w/o -2% through-plane strain. This figure counts the number of force constants values that appear in certain ranges among the finite number of force constants values in the systems.

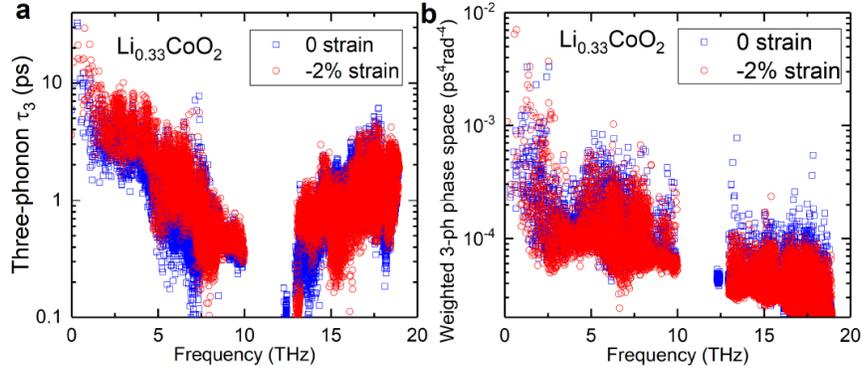

Fig. S35. The three-phonon relaxation times and (**b**) weighted three-phonon scattering phase spaces of $Li_{0.33}CoO_2$ w/o -2% through-plane strain.

## Section 18. Impact of lithium diffusion

The thermal conductivity derived from vibrational properties assumes that the atoms vibrate around fixed positions, i.e., their equilibrium positions, and do not flow around. It has, therefore, been a persistent question whether the diffusion of atoms makes a difference to the thermal conductivity of the materials with high ionic mobility[12]. To examine the impact of Li diffusion on the thermal conductivity of $Li_xCoO_2$, we estimate the mean time between two successive Li jumps. For 2D migrations, the ionic diffusion coefficient is defined as $D = d^2/4\Delta t$, where $d$ is a single jump distance, and $\Delta t$ is the time elapsed between two successive jumps. Providing the measured values[13] $D=$ 4-10 $\times 10^{-12}$ cm$^2$s$^{-1}$ and $d=$2.84 Å, the value of $\Delta t$ is estimated as 20-50 μs, which is several orders of magnitude longer than the phonon relaxation times (~ ps). In other words, during phonon relaxation time (1 ps), there are only 2-5 $\times 10^{-8}$ Li atoms that hop. If a hop is regarded as a "defect", the "defect" concentration is about 2-5 $\times 10^{-8}$, which has negligible impact on the vibrational and thermal properties of $Li_xCoO_2$. If an external electric field is applied, the "defect" concentration can be increased proportionally to the applied the electric field. However, to generate impact on the thermal transport, e.g., by making the "defect" concentration as 2~5%, the electric field is required to be as high as $10^6$ Vm$^{-1}$, which is beyond practice. Therefore, it is safe to conclude that the impact of Li diffusion on the thermal properties of Li$_x$CoO$_2$ is negligible.

## Section 19. Electronic thermal conductivity

Electronic contribution to thermal conductivity might also need to be considered for the thermal transport in batteries and memristors. Generally, electrical conductivity increases with decreasing $x$ since delithiation generates more holes. The electrical conductivity of Li$_x$CoO$_2$ and Li$_x$NbO$_2$, however, differs from samples to samples. Here, we take two available examples from the literature for discussion. In Ref.[14], the measured electrical conductivity increases from 0.1 to $10^3$ S·m$^{-1}$, when $x$ reduces from 1 to 0.6. Based on the Wiedemann–Franz law, the corresponding electronic thermal conductivity is $7\times10^{-7}$ ~ $7\times10^{-3}$ Wm$^{-1}$K$^{-1}$, which is negligible compared to the phonon contribution, 5.4 and 3.7 Wm$^{-1}$K$^{-1}$. In Ref.[15], the measured in-plane electrical conductivities of single-crystalline LiCoO$_2$ and Li$_{0.5}$CoO$_2$ are 20 and $4.65\times10^4$ S·m$^{-1}$. The corresponding thermal conductivities are $1.5\times10^{-4}$ and 0.34 Wm$^{-1}$K$^{-1}$, which are also much smaller than the lattice thermal conductivities. In Ref.[16], the measured electronic thermal conductivity is also negligible compared to the lattice one for Li$_x$NbO$_2$ ($0.3\leq x\leq1$). Therefore, electrical contribution to thermal transport may not be significant in many cases in practical devices.

## Section 20. Impact of porosity

Porosity might be an important origin for the low thermal conductivity of the practical LiCoO$_2$ cathodes as well. With the calculated thermal conductivity, $k_0$, of fully-dense LiCoO$_2$ in the preceding text, the impact of porosity can be readily predicted by using the effective medium approximation (EMA)[17–20] as $(1-P)^{3/2}\kappa_0$ or $(1-P)/(1+P/2)\kappa_0$, where $P$ is the porosity. For example, based on this model, we can estimate the LiCoO$_2$ thermal conductivity with $P$=54% is about 1.9 - 2.2 Wm$^{-1}$K$^{-1}$, which agrees well with experimental data[21] of 1.58 Wm$^{-1}$K$^{-1}$ for the same porosity, considering that other defects and grain size effect can further reduce the thermal conductivity in experiment.